\documentclass[12pt]{article}
\usepackage{draft,url} 
\usepackage[utf8]{inputenc}
\usepackage{hyperref}
\usepackage{graphicx,color,subfig,upgreek,mathtools}
\usepackage{amsfonts,amssymb,amsmath,multirow}
\usepackage{mathabx,epsfig}
\usepackage{pifont}
\usepackage[utf8]{inputenc}
\usepackage{makecell}
\usepackage{inputenc}
\usepackage{amsthm}
\usepackage{leftidx}
\usepackage{epstopdf}
\usepackage{booktabs}
\usepackage{pst-all}
\usepackage[off]{auto-pst-pdf}

\newtheorem{theorem}{Theorem}
\newtheorem{lemma}[theorem]{Lemma}

\usepackage{tikz-cd}
 
\newcommand{\Z}{\mathbb Z}
\def\U{\mathrm{U}(1)}
\def\L{\mathcal{L}}
\def\NS{\mathrm{NS}}
\def\R{\mathrm{R}}

 	\def\C{\mathcal{C}}
	\def\exC{\Breve{\mathcal{C}}}

\begin{document}

\begin{titlepage}

\title{Symmetry-preserving boundary of (2+1)D fractional quantum Hall states}

\author{Ryohei Kobayashi
}

\address{Condensed Matter Theory Center and Joint Quantum Institute, Department of Physics, University of Maryland, College Park, Maryland 20472 USA
 }

\abstract{We investigate symmetry-preserving gapped boundary of (2+1)D topological phases with global symmetry, which can be either bosonic or fermionic. We develop a general algebraic description for gapped boundary condition for symmetry-enriched or fermionic topological phases, extending the framework of Lagrangian algebra anyon for bosonic phases without symmetry. We then focus on application to the case with U(1) symmetry. We derive new obstructions to symmetry-preserving gapped boundary for  U(1)$^f$-symmetric (2+1)D fermionic topological phases, which are beyond chiral central charge $c_-$ and electric Hall conductivity $\sigma_H$. These obstructions are given by a simple Gauss-Milgram type formula valid for super-modular category, and regarded as a higher version of $c_-$ and $\sigma_H$.}

\end{titlepage}

\eject
\tableofcontents

\unitlength = .8mm
\mbox{}\\

\section{Introduction}

Fractional quantum Hall (FQH) states are the most well-studied class of topological phases, since their experimental discovery in 1982~\cite{tsui1982}. A FQH state hosts a topological order in (2+1)D with U(1) global symmetry, characterized by the existence of fractionally charged quasiparticles called anyons~\cite{laughlin1983, wilczek1982}. 

The boundary of a FQH state has a gapless edge state, protected by non-zero quantized electric Hall conductivity $\sigma_H$ in the bulk. This gapless edge state cannot be gapped out while preserving U(1) global symmetry. This protection of gapless edge mode is understood as a consequence of perturbative U(1) anomaly on (1+1)D boundary theory present when $\sigma_H\neq 0$, which cannot be matched by gapped degrees of freedom.

Even in the absence of global symmetry, the boundary of a (2+1)D topological ordered state is enforced to be gapless when its chiral central charge $c_-$ is nonzero. This is the most fundamental obstruction to having a gapped edge state, and understood as a result of gravitational anomaly on the (1+1)D boundary theory present when $c_-\neq 0$. For example, the Moore-Read state~\cite{moore1991, Banerjee2018}, the most well-known example of a non-Abelian FQH state, carries $c_-=3/2$, and then its boundary cannot be gapped out even when we forget about U(1) global symmetry.

The electric and thermal Hall conductivity $\sigma_H$ and $c_-$ give important obstructions to a gapped edge state of (2+1)D topological order, but do not cover complete obstructions to a gapped edge state.
Actually, even in the absence of global symmetry where we do not have $\sigma_H$, the condition $c_-=0$ alone does not guarantee the existence of a gapped boundary. For example, it is known that $\U_2\times \U_{-4}$ Chern-Simons theory does not admit a gapped boundary condition, though this theory carries $c_-=0$. 
The gapped boundary condition of a bosonic topological phase in (2+1)D without global symmetry has been extensively studied in~\cite{Kitaev:2011dxc, Kapustin:2010hk, Kapustin:2013nva, Barkeshli:2013jaa, Hung:2014tba, Wang:2012am, Lan:2014uaa, Kong:2019byq}, and algebraically characterized by the Lagrangian algebra anyon of a modular tensor category~\cite{kaidi2021higher, davydov2013witt, Fuchs:2012dt, davydov2013structure}. 
Physically, the Lagrangian algebra represents the set of condensed anyons on the boundary, where a gapped boundary is generally obtained by performing anyon condensation~\cite{Bais:2008ni, Burnell:2017otf}.
The obstruction to the gapped boundary is most generally understood as the absence of the Lagrangian algebra.

Though it is not practically easy to see if a given general bosonic topological order admits the Lagrangian algebra or not, it is known that the obstructions to gapped boundary can be partially captured by easily computable quantities called higher central charge $\xi_n$ labeled by a positive integer $n$~\cite{Ng2018higher, Ng2020higher}. That is, for a given bosonic (2+1)D topologically ordered state, one can compute these obstructions to gapped boundary by a simple formula in terms of the properties of anyons,
\begin{align}
    \xi_n = \frac{\sum_{a} d_a^2\theta_a^n}{|\sum_{a} d_a^2\theta_a^n|}
\end{align}
where the sum is over all anyons in the topological order. $d_a$ is quantum dimension, and $\theta_a$ is topological twist (i.e., self-statistics) of an anyon $a$.
When $n=1$, $\xi_1$ gives chiral central charge of the bosonic topological phase modulo 8~\cite{Kitaevanyons},
\begin{align}
    \xi_1 = \frac{\sum_{a} d_a^2\theta_a}{|\sum_{a} d_a^2\theta_a|}= e^{\frac{2\pi i}{8}c_-},
    \label{eq:gauss}
\end{align}
$\{\xi_n\}$ hence provide higher generalizations of the chiral central charge $c_-$. 
In general, one can only obtain $c_-$ mod 8 for a given data of the anyons. This is because there exists a (2+1)D bosonic invertible phase called $E_8$ state which does not carry anyons and has $c_-=8$~\cite{KitaevE8}. Hence, one can shift $c_-$ by integer multiple of 8 by stacking a copy of $E_8$ phases on the topological order, without changing the data of anyons. Regarding this ambiguity by $8\Z$, the formula~\eqref{eq:gauss} completely determines $c_-$ mod 8 using the properties of anyons. The formula~\eqref{eq:gauss} is sometimes called the Gauss-Milgram formula.

Not all $\xi_n$ for $n\in\Z$ correspond to obstructions to gapped boundary. To be precise,~\cite{kaidi2021higher} proved that $\xi_n=1$ for all $n$ such that $\gcd(n, N_{\mathrm{FS}})=1$ give necessary conditions for admitting a gapped boundary. Here, $N_{\mathrm{FS}}$ is called the Frobenius-Schur exponent, which is defined as a smallest positive integer such that $\theta_a^{N_{\mathrm{FS}}}=1$ for all anyons $a$. 
These quantities $\{\xi_n\}$ for $n>1$ provide obstructions to gapped boundary beyond $c_-$. For example, we have $\xi_3=-1$ for $\U_2\times \U_{-4}$ Chern-Simons theory, which shows that $\U_2\times \U_{-4}$ does not admit a gapped boundary even though $c_-=0$.

In the presence of global symmetry, one can ask if a given (2+1)D topological order admits a gapped boundary preserving the symmetry. 
In this paper, we investigate  obstructions to gapped boundaries in (2+1)D topological phases with global symmetry, for both bosonic and fermionic phases.
For bosonic topological phases, we show that the global symmetry puts additional constraints on the Lagrangian algebra required for preserving global symmetry, which gives rise to further obstructions to gapped boundary. For example, the electric Hall conductivity $\sigma_H$ is regarded as such an additional obstruction that arises by enriching the phase with $\U$ global symmetry. 

For fermionic topological phases, the algebraic formulation of gapped boundary in terms of the Lagrangian algebra has not been developed yet, even in the absence of the global symmetry. We first extend the formalism of the Lagrangian algebra valid for fermionic topological phases, and then study its symmetry enrichment mainly focusing on the U(1)$^f$ symmetry, i.e., U(1) symmetry charging fermions. In particular, we derive new obstructions to symmetry-preserving gapped boundary beyond $\sigma_H$ and $c_-$, for a fermionic phase with U(1)$^f$ symmetry. 

\subsection{Summary of results}

Here we summarize the results of the paper. 
First, we study the (2+1)D bosonic topological phases with global symmetry $G$. 
In general, global symmetry of a (2+1)D topological ordered phase is characterized by symmetry fractionalization on anyons~\cite{barkeshli2019}. 
Roughly speaking, the symmetry fractionalization means that the global symmetry acts projectively on anyons. An example of symmetry fractionalization is found in FQH states where anyons carry fractional charge under U(1) symmetry. 

Then, we derive several constraints on the symmetry fractionalization data of anyons required for the existence of symmetry-preserving gapped boundary.
We basically show that symmetry fractionalization data for Lagrangian algebra anyons (i.e., condensed anyons on the boundary) must be trivial, in order to realize a symmetry-preserving gapped boundary. For example, in the case of U(1) symmetry, we show that the Lagrangian algebra anyons must carry trivial fractional charge. 

Based on the constraints on symmetry fractionalization data of the Lagrangian algebra anyons, we can generally show that some specific set of anyons must be condensed to realize a symmetry-preserving gapped boundary. For example, in the case of U(1) symmetry, a special anyon $v$ called a vison must be condensed. This in particular means that $v$ must be a boson, $\theta_v=1$. Here, it is known that $\theta_v$ computes the electric Hall conductivity as~\cite{lapa2019, benini2019, kobayashi2021spinc}
\begin{align}
    \theta_v=e^{i \pi {\sigma}_H},
\end{align}
where we define the Hall conductivity by the electromagnetic response locally given by the Chern-Simons action $-\frac{\sigma_H}{4\pi}A\mathrm{d}A$.

So, the constraint $\theta_v=1$ corresponds to vanishing electric Hall conductivity.
As an application, we show that in (2+1)D bosonic Abelian topological phases with U(1) symmetry, $\theta_v=1$ together with $\xi_n=1$ for all $n$ such that  $\gcd(n,\frac{N_{\mathrm{FS}}}{\gcd(n,N_{\mathrm{FS}} )})=1$ give necessary and sufficient conditions for symmetry-preserving gapped boundary. This generalizes the result in~\cite{kaidi2021higher} for bosonic Abelian topological phases to the case with U(1) global symmetry.

Next, we investigate gapped boundary condition for (2+1)D fermionic topological phases.
We first develop a general algebraic framework for gapped boundary of fermionic topological phases, by extending the formalism of the Lagrangian algebra anyon for bosonic phases to fermionic cases.~\footnote{In this paper, we restrict ourselves to fermionic phases that no anyons $\sigma$ carrying vortex of $\Z_2^f$ fermion parity symmetry satisfy $\sigma\times \psi = \sigma$, where $\psi$ is a fermion physically regarded as an electron. When we are interested in fermionic phases with $\U^f$ symmetry which is main interest in this paper, there exists no such anyons since the assignment of U(1) charge on anyons cannot be consistent with the fusion rule $\sigma\times \psi = \sigma$. So, this assumption does not lose generality in that case.}
We also make a generalization of the formalism to the case of symmetry-preserving gapped boundary, mainly focusing on the case with $\U^f$ symmetry.

As an application of the above formalism for fermionic phases, we propose new obstructions to symmetry-preserving gapped boundary of (2+1)D fermionic topological phases with $\U^f$ symmetry, given by
\begin{align}
    \zeta_n:=\frac{\sum_{a\in\mathcal{C}}e^{i\pi Q_a} d_a^2\theta_a^n}{|\sum_{a\in\mathcal{C}}e^{i\pi Q_a} d_a^2\theta_a^n|}
\end{align}
where $Q_a$ is fractional charge of an anyon $a$, and $\C$ is a super-modular tensor category that characterizes the properties of anyons in a fermionic phase.
We show that $\zeta_n=1$ for all $n$ such that $\gcd(n,N_{\mathrm{FS}})=1$ give necessary conditions to symmetry-preserving gapped boundary.

The above quantities $\{\zeta_n\}$ are regarded as higher versions of $c_-$ and $\sigma_H$, in the sense that $\{\zeta_n\}$ provide obstructions to symmetry-preserving gapped boundaries beyond $c_-$ and $\sigma_H$. For $\U^f$-symmetric fermionic topological phases, $c_-$ and $\sigma_H$ are given by~\cite{kobayashi2021spinc}
\begin{align}
    e^{-2\pi i c_-}= (\sqrt{2}\mathcal{D})^8\cdot \frac{\sum_{a\in\mathcal{C}}e^{3i\pi Q_a} d_a^2\theta_a}{\left(\sum_{a\in\mathcal{C}}e^{i\pi Q_a} d_a^2\theta_a\right)^9}
    \label{eq:c-spinc}
\end{align}
\begin{align}
    e^{-2\pi i {\sigma}_H}= \frac{\sum_{a\in\mathcal{C}}e^{3i\pi Q_a} d_a^2\theta_a}{\sum_{a\in\mathcal{C}}e^{i\pi Q_a} d_a^2\theta_a}
    \label{eq:sigmaspinc}
\end{align}
and $\zeta_n$ cannot be expressed by any multiplication of $e^{-2\pi i c_-}$ and $e^{-2\pi i \sigma_H}$.
As we can see in~\eqref{eq:c-spinc} and~\eqref{eq:sigmaspinc}, one can only obtain $c_-$ and $\sigma_H$ mod 1 for a given data of super-modular category. This is because there exists a (2+1)D fermionic invertible phase characterized by Spin$^c$ Chern-Simons theory characterized by the theta term~\cite{Seiberg2016Gapped}
\begin{align}
    \int\left(\frac{2\pi}{192\pi^2}\mathrm{Tr}(R\wedge R)-\frac{2\pi}{8\pi^2}F\wedge F\right)
\end{align}
which carries $c_-=\sigma_H=1$.

\subsection{Organization of paper}
This paper is organized as follows. In Sec.~\ref{sec:bosonic}, after reviewing an algebraic formalism of gapped boundary of bosonic phases called the Lagrangian algebra anyon, we make a generalization of the framework to symmetry-enriched topological phases. We then derive a necessary and sufficient condition for U(1)-preserving gapped boundary in the case of bosonic Abelian topological order with U(1) symmetry.
In Sec.~\ref{sec:fermionic}, we move to fermionic topological phases, and describe a generic theory for anyon condensation in fermionic non-Abelian topological order. After studying various properties of fermionic gapped boundary, we discuss the case with $\U^f$ symmetry.
In Sec.~\ref{sec:higher}, we derive obstructions to $\U^f$-preserving gapped boundary of fermionic phases with $\U^f$ symmetry. Review of concepts used in this paper and detailed calculations are relegated to appendices.

\section{Gapped boundary of bosonic topological phases}
\label{sec:bosonic}
\subsection{Review: Lagrangian algebra anyons}
We begin with a brief review for some properties of gapped boundary of a bosonic topological phase without global symmetry. See~\cite{kaidi2021higher} for detailed descriptions.

\begin{figure}[htb]
\centering
\includegraphics{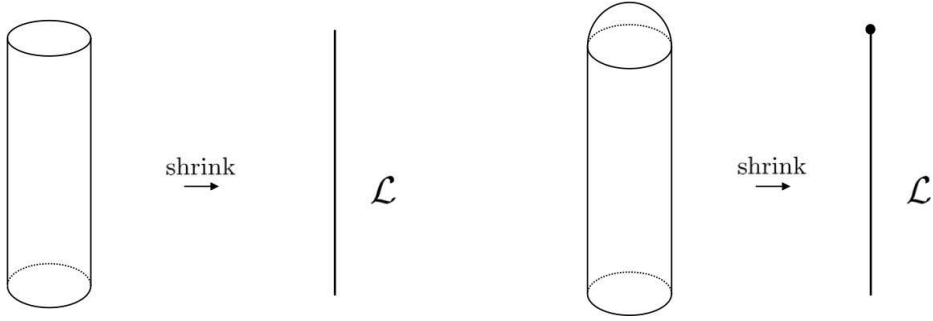}
\caption{One can prepare a cylinder of a gapped boundary, then shrinking it results in a line operator. One can cap off the cylinder by a gapped boundary, so the line operator can end at a point, which means that $\mathrm{Hom}(\mathcal{L},1)$ is not empty.}
\label{fig:shrink}
\end{figure}

Gapped boundary of a (2+1)D bosonic topological quantum field theory (TQFT) is algebraically described by an object called a Lagrangian algebra anyon~\cite{kaidi2021higher, davydov2013witt, Fuchs:2012dt, davydov2013structure}. The idea is that when a TQFT admits a topological gapped boundary condition, we consider cutting out a solid cylinder from a spacetime 3-manifold. We introduce a gapped boundary condition on the boundary of the resulting manifold, getting a cylinder of gapped boundary.
We shrink the radius of the cylinder of a gapped boundary, then it eventually becomes a topological line operator of a TQFT, see Fig.~\ref{fig:shrink}. So, the tube of the gapped boundary after shrinking is expressed as a sum of simple anyons in a modular tensor category $\C$,
\begin{align}
    \mathcal{L}=\bigoplus_{a\in\C} Z_{0a} a,
\end{align}
with non-negative integers $Z_{0a}$. This object $\mathcal{L}$ is called a Lagrangian algebra anyon. Since one can cap off the tube on the top of it and introduce gapped boundary on the cap, the tube of a gapped boundary can end at a point. It means that $\mathrm{Hom}(\mathcal{L},1)$ is not empty, and hence $Z_{00}>0$. See Fig.~\ref{fig:shrink}. In general, the Lagrangian algebra anyon with $Z_{00}>1$ is known to decompose into the sum of those with $Z_{00}=1$. The simple gapped boundary condition is hence described by the Lagrangian algebra anyon satisfying $Z_{00}=1$.

When $Z_{0a}>0$ for some anyon $a\in\C$, $\mathrm{Hom}(\mathcal{L}\times a,\mathcal{L})$ is not empty since $Z_{00} > 0$. This implies that the Wilson line of $a$ can end on the tube of gapped boundary, meaning that $a$ is condensed on the boundary. So, anyons with $Z_{0a} > 0$ is physically regarded as a set of condensed anyons. 

$\mathcal{L}$ satisfies nice properties under modular $S, T$ transformations. The vector $\{Z_{0a}\}$ turns out to be an eigenvector of modular $S$ and $T$ matrices:
\begin{align}
    \sum_{b\in\C} S_{ab} Z_{0b} = Z_{0a}, \quad \sum_{b\in\C} T_{ab} Z_{0b} = Z_{0a}
    \label{eq:STZ}
\end{align}
Since $S$ and $T$ of modular tensor category $(ST)^3=e^{\frac{2\pi i}{8}c_-}S^2$, the existence of the Lagrangian algebra anyon with~\eqref{eq:STZ} implies $c_-=0$ mod $8$.

We can consider a fusion space of the Lagrangian algebra anyon $V^{\mathcal{L}\mathcal{L}}_{\mathcal{L}}$ by taking a junction of three tubes of gapped boundary. We can then talk about the $F$- and $R$-move of tubes with junctions, which turn out to be trivial:
\begin{align}
    (F^{{\mathcal{L}}{\mathcal{L}}{\mathcal{L}}}_{{\mathcal{L}}})_{{\mathcal{L}}, {\mathcal{L}}}\cdot \ket{{\mu}}\otimes  \ket{{\mu}} = \ket{{\mu}}\otimes \ket{{\mu}}
    \label{eq:lagrangianFboson}
\end{align}
\begin{align}
    R^{{\mathcal{L}}{\mathcal{L}}}_{\mathcal{L}}\ket{\mu}=\ket{\mu}
    \label{eq:lagrangianRboson}
\end{align}

\subsection{Symmetry-preserving gapped boundary}
\label{sec:symmetry}
Here, we consider a (2+1)D TQFT with global symmetry $G$, and study gapped boundary condition that preserves the global symmetry of the bulk.
We assume that the background gauge field is realized by a network of codimension-1 symmetry defects. A symmetry defect in the bulk can transversally end on the gapped boundary at a line, and the ending line then defines a symmetry defect of the gapped boundary theory.

\begin{figure}[htb]
\centering
\includegraphics{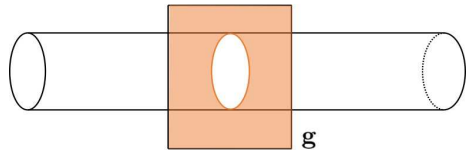}
\caption{A tube of gapped boundary passes through a symmetry defect.}
\label{fig:crossingtube}
\end{figure}

First, we can cut out a tube from a 3-manifold and introduce the gapped boundary condition on the boundary. We then consider a symmetry defect across the carved tube, see Fig.~\ref{fig:crossingtube}. In this setup, since the boundary condition should be invariant under the symmetry action, we require that the symmetry action leaves the Lagrangian algebra anyon invariant,
\begin{align}
   Z_{0a} = Z_{0, \rho_{\mathbf{g}}(a)} \quad \text{for $\mathbf{g}\in G$}.
   \label{eq:labelpreserve}
\end{align}

To derive a further constraint on the Lagrangian algebra anyon, we argue that gauge transformations of flat background $G$ gauge field in the bulk-boundary system leave the partition function invariant. We verify this statement based on an similar argument to Ref.~\cite{Thorngren2021end}, which showed that a theory must be free of 't Hooft anomaly if it admits a symmetry-preserving boundary condition. Following Ref.~\cite{Thorngren2021end}, we put several axioms about the properties of background gauge transformations in the bosonic phases:

\begin{enumerate}
    \item The partition function is invariant under smooth isotopies of symmetry defects.
    \item The partition function is invariant under introducing or removing spherical components of the symmetry defects whose interior does not contain any defects or other operator insertions.
    \item We can perform a recombination of the symmetry defects by Pachner moves on either bulk or boundary, and it only has the effect of shifting the partition function by a phase.
\end{enumerate}

\begin{figure}[htb]
\centering
\includegraphics{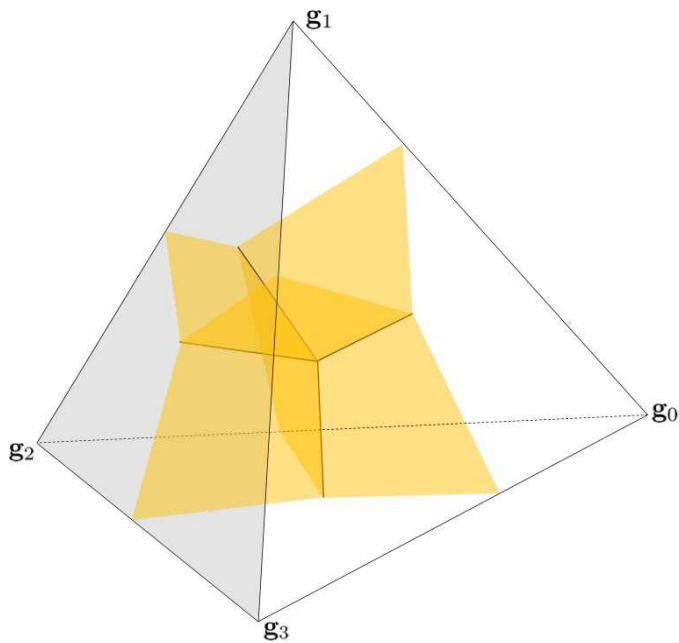}
\caption{Poincar\'e dual of background $G$-gauge field on a 3-simplex is represented by yellow sheets, which are regarded as codimension-1 symmetry defects. Background gauge field on a 3-simplex is characterized by group elements $\mathbf{g}_i$ assigned on each 0-simplex $i$. That is, a defect on the dual of a 1-simplex $\langle ij\rangle$ is given by $\mathbf{g}_i^{-1}\mathbf{g}_j$. A 3-sphere $S^3$ consists of five 3-simplices regarded as a boundary of a 4-simplex, and a bubble of a symmetry defect on $S^3$ is constructed by connecting up the symmetry defects on each 3-simplex along 2-simplices. }
\label{fig:3simplex}
\end{figure}

Then, we think of creating a bubble of $G$ symmetry defects in 3D bulk given in a following way. First, consider a 3-sphere $S^3$ composed of five 3-simplices, regarded as a boundary of a 4-simplex. The bubble of symmetry defects is given by the Poincar\'e dual of flat background $G$-gauge field on the triangulated $S^3$ (see Fig.~\ref{fig:3simplex}), stereographically projected onto a 3-ball $D^3$. It can be checked that this bubble can be eliminated from the bulk by performing a single Pachner move in 3D, up to other moves pushing the diagram into the boundary without producing a phase~\cite{Thorngren2021end}. 

The phase produced by a Pachner move is expressed as a function  $\omega: G^5\to \U$ where $G^5$ denotes group elements on  five 0-simplices on $S^3$ that characterizes a background gauge fields on $S^3$. $\omega$ turns out to take a value in a 4-cocycle $Z^4(BG, \U)$, and characterizes a 't Hooft anomaly of the (2+1)D bulk~\cite{chen2012spt}. Then, it has been shown in~\cite{Thorngren2021end} that the 't Hooft anomaly must be trivial when the bulk admits a symmetry-preserving gapped boundary, so we set $\omega=0$. It means that the bubble can be created without producing any phase.

\begin{figure}[htb]
\centering
\includegraphics{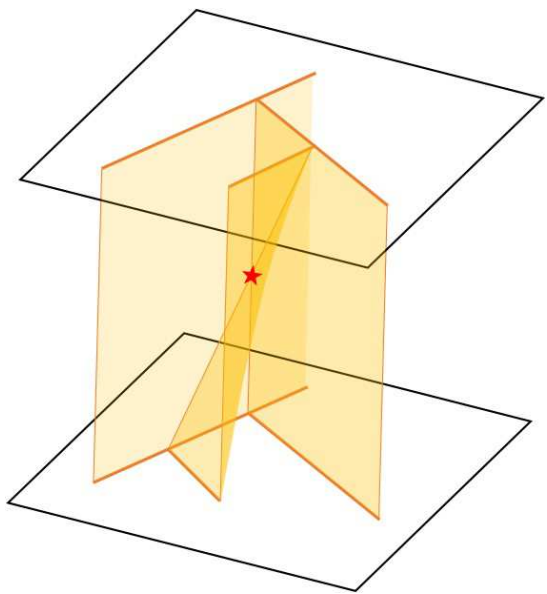}
\caption{At the center of a 3-simplex, there is a codimension-3 junction of symmetry defects (a red starred point). When this junction gets absorbed by the boundary, it causes the Pachner move of symmetry defects on the boundary.}
\label{fig:pachner}
\end{figure}

Then, we eliminate the bubble by pushing it into the boundary. It involves a Pachner move on boundary, every time a codimension-3 junction of defects at the center of a 3-simplex gets absorbed in the boundary, see Fig.~\ref{fig:pachner}. We write the phase produced by a boundary Pachner move as $\alpha: G^4\to \U$, as a function of group labels on the four vertices of a 3-simplex. The overall phase factor to eliminate the bubble is given by $\delta\alpha$ evaluated on a 4-simplex. 

Since $\delta\alpha=0$ due to the vanishing of 't Hooft anomaly in the bulk, $\alpha$ takes its value in $Z^3(BG,\U)$. 
When the gauge transformation on the boundary shifts the partition function by a phase $\alpha\in Z^3(BG,\mathrm{U}(1))$, one can cancel the phase ambiguity by coupling the bulk with a (2+1)D bosonic SPT phase characterized by the 3-cocycle $\alpha$. This process makes the bulk-boundary system gauge invariant without modifying the boundary condition, or the data of symmetry-enriched topological phase in the bulk (i.e., anyons and their symmetry fractionalization). Hence, one can assume without loss of generality that gauge transformations of flat background $G$ gauge field in the bulk-boundary system leave the partition function invariant.

\begin{figure}[htb]
\centering
\includegraphics{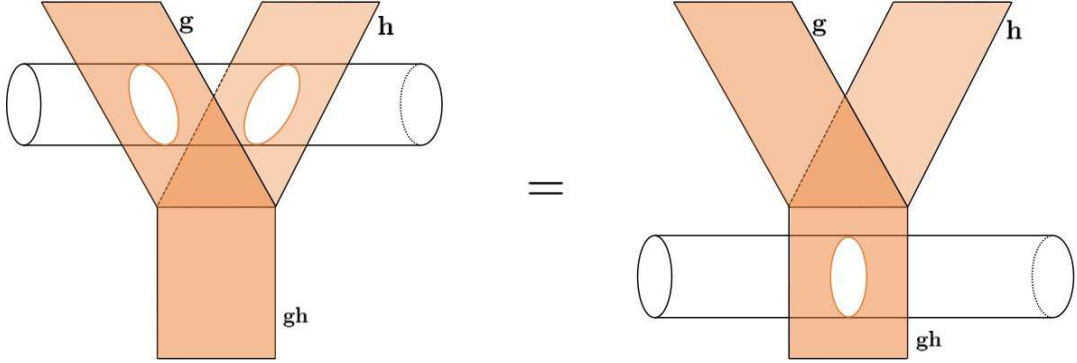}
\caption{A junction of symmetry defects can pass through a tube of gapped boundary by background gauge transformation, realized by combination of a couple of Pachner moves together with introducing a trivial spherical bubble of a symmetry defect on the boundary.}
\label{fig:etamove}
\end{figure}

Now, let us consider a junction of three symmetry defects $\mathbf{g}, \mathbf{h}, \mathbf{gh}\in G$ in the 3-manifold, then carve out a tube piercing the symmetry defects, as shown in Fig.~\ref{fig:etamove}. We introduce the gapped boundary condition on the boundary of the tube, where symmetry defects in 3D end on symmetry defects on the boundary theory. 

Then, by performing background gauge transformation, one can move the junction of symmetry defects across the tube. By shrinking the tube during the process of the gauge transformations, one can see that symmetry fractionalization on the Lagrangian algebra anyon is trivial (see Appendix~\ref{app:symfrac} for a review of symmetry fractionalization),
\begin{align}
    \eta_{\mathcal{L}}(\mathbf{g},\mathbf{h})=1.
    \label{eq:etalagrangian}
\end{align}

\begin{figure}[htb]
\centering
\includegraphics{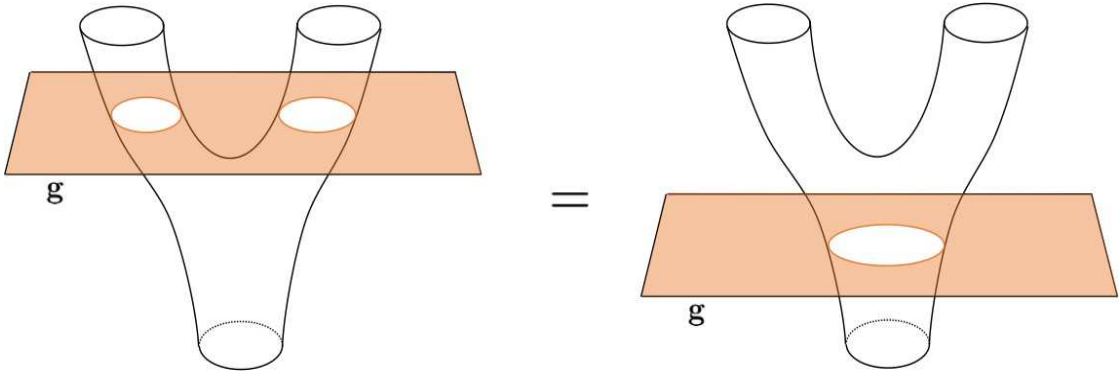}
\caption{A symmetry defect $\mathbf{g}\in G$ can pass through a junction of gapped boundaries by background gauge transformation, realized by a single Pachner move on the boundary.}
\label{fig:Umove}
\end{figure}

Also, consider cutting out the junction of three tubes and introducing the gapped boundary condition on its boundary, see Fig.~\ref{fig:Umove}. Then, by performing background gauge transformation across the junction, we obtain
\begin{align}
    U_{\mathbf{g}}(\mathcal{L},\mathcal{L}; \mathcal{L})=1,
    \label{eq:Ulagrangian}
\end{align}
which means that $G$ symmetry acts trivially on the fusion space $V^{\mathcal{L},\mathcal{L}}_{\mathcal{L}}$.

In particular, suppose that we can fix a gauge as $U_{\mathbf{g}}(a,b;c)=1$ for any $a,b,c\in\C$ with $N^{ab}_c>0$ and $\mathbf{g}\in G$, and consider a symmetry-preserving gapped boundary in this fixed gauge. We can then express symmetry fractionalization as $\eta_a(\mathbf{g},\mathbf{h})=M_{a, \mathfrak{t}(\mathbf{g},\mathbf{h})}$ using some Abelian anyon $\mathfrak{t}(\mathbf{g},\mathbf{h})\in\mathcal{C}$, which gives an element of $Z^2_{\rho}(BG,\mathcal{B})$ where $\mathcal{B}$ is a group of Abelian anyons in $\C$. Then, by using the property $SZ=Z$ of the Lagrangian algebra anyon we have
\begin{align}
    \sum_{b\in\mathcal{A}}S_{\mathfrak{t}(\mathbf{g},\mathbf{h}),b} Z_{0b} = Z_{0, \mathfrak{t}(\mathbf{g},\mathbf{h})}
    \label{eq:SZZfort}
\end{align}
where $\mathcal{A}$ is the set of anyons with $Z_{0a}>0$. Since $S_{\mathfrak{t}(\mathbf{g},\mathbf{h}),b} = \frac{d_b M^*_{\mathfrak{t}(\mathbf{g},\mathbf{h}),b}}{\mathcal{D}}$ and $M_{b, \mathfrak{t}(\mathbf{g},\mathbf{h})}=\eta_b(\mathbf{g},\mathbf{h})=1$ when $b\in\mathcal{A}$,~\eqref{eq:SZZfort} is rewritten as
\begin{align}
    Z_{0, \mathfrak{t}(\mathbf{g},\mathbf{h})}=\frac{1}{\mathcal{D}}\sum_{b\in\mathcal{A}} d_b Z_{0b}=Z_{00}>0,
\end{align}
so we have $Z_{0, \mathfrak{t}(\mathbf{g},\mathbf{h})}>0$. Due to $TZ=Z$, we can see that $\mathfrak{t}(\mathbf{g},\mathbf{h})$ must be a boson, $\theta_{\mathfrak{t}(\mathbf{g},\mathbf{h})}=1$. 
More precisely, for a given symmetry fractionalization class $[\mathfrak{t}]\in H_{\rho}^2(BG,\mathcal{B})$,~\eqref{eq:etalagrangian} should be understood as requiring the existence of natural isomorphism that picks a specific cocycle representative of $[\mathfrak{t}]$, so that $\eta_a(\mathbf{g},\mathbf{h})=1$ for $a\in\mathcal{A}$. The statement $\theta_{\mathfrak{t}(\mathbf{g},\mathbf{h})}=1$ is valid after performing such a natural isomorphism so that $\eta$ satisfies~\eqref{eq:etalagrangian}. 

Here we comment on the simplicity of the Lagrangian algebra anyon in the presence of the global symmetry.
Though a gapped boundary condition with $Z_{00} > 1$ always decomposes into simple ones with $Z_{00}=1$ in the absence of the global symmetry, the symmetry-preserving boundary with $Z_{00} > 0$ does not necessarily decompose into simple symmetric ones. In fact, for a given simple symmetry-breaking gapped boundary that violates~\eqref{eq:labelpreserve}, one can sum over the orbit of $G$ symmetry actions on it to obtain a $G$-symmetric boundary condition, which does not decompose into symmetry-preserving ones in general.
However, when $G$ does not permute the anyons where~\eqref{eq:labelpreserve} is satisfied, the constraints~\eqref{eq:etalagrangian},~\eqref{eq:Ulagrangian} are obviously preserved for each decomposed component of the Lagrangian algebra anyons. Hence, we conjecture that the symmetric boundary condition with $Z_{00}>1$ always decomposes into simple symmetric ones with $Z_{00}=1$ when $[\rho]$ is trivial. 

\subsection{Example: $G=\U$}
When $G=\U$, the symmetry does not permute anyons. Symmetry fractionalization is then characterized by an assignment of fractional charge defined as follows. 
    For a given set of $\{\eta, U\}$, one can define fractional charges of anyons that characterize the U(1) symmetry fractionalization as follows. For a fixed anyon $a$, let $n$ be the smallest integer such that $a^n$ contains the identity as a fusion product. Choose a sequence of anyons $a, a^2,\dots a^{n}=1$ such that $a\times a^k$ contains $a^{k+1}$ as a fusion product. Then define a fractional charge $Q_a\in\mathbb{R}/\Z$ as \cite{bulmashSymmFrac}
\begin{align}
    e^{2\pi i Q_a}:= \prod_{m=1}^{n-1}\eta_a\left(\frac{1}{n}, \frac{m}{n}\right)U_{\frac{1}{n}}(a,a^m; a^{m+1}),
\end{align}
where the elements of $\U=\mathbb{R}/\Z$ is labeled by the numbers in $[0,1)$. One can check that the quantity $e^{2\pi i Q_a}$ is gauge-invariant. Since the map $[\rho]$ is trivial, we can fix a gauge where $U = 1$,~\footnote{For modular tensor categories with the trivial symmetry action $[\rho]$, the existence of the gauge with $U=1$ is rigorously shown in Ref.~\cite{benini2019} when all fusion coefficients satisfy $N^{c}_{a,b}\in\{0,1\}$. For the case with generic fusion coefficients, it is a conjecture given in Ref.~\cite{barkeshli2019}.} so that $\eta$ satisfies ${\eta_c({\bf g}, {\bf h})}=\eta_a({\bf g}, {\bf h})\eta_b({\bf g}, {\bf h})$ when $N^{c}_{ab}>0$.
One can then see that
\begin{align}
    e^{2\pi i Q_a}e^{2\pi i Q_b}=e^{2\pi i Q_c}\quad \text{when $N_{a,b}^c\neq 0$.} \label{eq:fusionu1charge}
\end{align}

In the gauge where $U = 1$, we can write the phases $\eta_a({\bf g}, {\bf h})$ as
\begin{align}
    \eta_a({\bf g},{\bf h}) = M_{a, \mathfrak{t}({\bf g},{\bf h})} ,
\end{align}
for Abelian anyon $\mathfrak{t}\in Z^2(B\U, \mathcal{B})$ where $\mathcal{B}$ is the set of Abelian anyons in $\mathcal{C}$. A representative 2-cocycle $\mathfrak{t}$ is given by $\mathfrak{t}({\bf g}, {\bf h}) = v^{{\bf g} + {\bf h} - [ {\bf g} + {\bf h}]} ,$
where $v \in \mathcal{B}$ is referred to as the vison, $\mathbf{g},\mathbf{h}\in \mathbb{R}/\Z$ takes the values in $[0,1)$, and $[\mathbf{g}+\mathbf{h}]$ means the sum mod 1. 
Using the vison $v$, the fractional charge can be rewritten as \cite{cheng2016lsm}
\begin{align}
    e^{2\pi i Q_a} = M_{a,v}.
    \label{eq:vison}
\end{align}

Then, suppose that a bosonic TQFT with U(1) symmetry admits a U(1) symmetry-preserving gapped boundary, after setting the gauge $U=1$ by a suitable natural isomorphism.
Then,~\eqref{eq:etalagrangian} implies that 
\begin{align}
    e^{2\pi i Q_a}= 1 \quad \text{when $a\in\mathcal{A}$.}
\end{align}
Hence, the vison $v$ has trivial mutual braiding with all $a\in\mathcal{A}$, $M_{a,v}=1$. 
Using the property of Lagrangian algebra anyon $SZ=Z$ and $S_{v,b} = \frac{d_b M^*_{v,b}}{\mathcal{D}}$, we obtain
\begin{align}
    Z_{0v}= \sum_{b\in \mathcal{A}} S_{v,b}Z_{0b} = \frac{1}{\mathcal{D}}\sum_{b\in \mathcal{A}}d_bZ_{0b}=Z_{00},
\end{align}
so we have $Z_{0v}>0$.
According to $TZ=Z$, we immediately see that the vison $v$ must be a boson, $\theta_v=1$.

\subsection{Application: bosonic Abelian TQFT with U(1) symmetry}
As a straightforward application, we study an Abelian (2+1)D bosonic TQFT with U(1) symmetry. We prove the following statement:
\begin{theorem}
A $c_-=0$ bosonic Abelian TQFT with U(1) symmetry has a U(1) symmetric gapped boundary if and only if the vison $v$ in~\eqref{eq:vison} has $\theta_v=1$, and the higher central charge $\xi_n$ defined as
\begin{align}
    \xi_n:=\frac{\sum_{a\in\mathcal{C}}\theta_a^n}{|\sum_{a\in\mathcal{C}}\theta_a^n|}
\end{align}
becomes 1 for all $n$ such that $\gcd(n,\frac{N_{\mathrm{FS}}}{\gcd(n,N_{\mathrm{FS}} )})=1$. 
\end{theorem}
Here, $N_{\mathrm{FS}}$ is the Frobenius-Schur exponent defined as the smallest positive integer such that $\theta_a^{N_{\mathrm{FS}}}=1$ for all $a\in\mathcal{C}$. 
\begin{proof}
It was shown in~\cite{kaidi2021higher} that the TQFT has a gapped boundary only if $\xi_n=1$ for all $n$ with $\gcd(n,\frac{N_{\mathrm{FS}}}{\gcd(n,N_{\mathrm{FS}} )})=1$, and we have shown that $\theta_v=1$ is necessary for symmetry-preserving gapped boundary. So the ``only if'' part is done. We prove the ``if'' part. To see this, we use the lemma shown in~\cite{kaidi2021higher}.
\begin{lemma}
An Abelian TQFT $\mathcal{C}$ with Frobenius-Schur exponent $N_{\mathrm{FS}}$ admits a factorization
\begin{align}
    \mathcal{C}=\mathcal{C}_{p_1}\times \mathcal{C}_{p_2}\times \dots \times \mathcal{C}_{p_k}
\end{align}
where $\mathcal{C}_{p_i}$ are TQFTs labelled by distinct primes $p_i$, such that the number of anyons in $\mathcal{C}_{p_i}$
is a positive integer power of $p_i$. If we denote the Frobenius-Schur exponents of $\mathcal{C}_{p_i}$ by $N_i$,
then $N_{\mathrm{FS}}=N_1N_2\dots N_k$. \quad
$\qedsymbol$
\end{lemma}
Assume that $\xi_n=1$ for all $n$ such that $\gcd(n,\frac{N_{\mathrm{FS}}}{\gcd(n,N_{\mathrm{FS}} )})=1$. This implies that $\xi_n(\mathcal{C}_{p_r})=1$ for all $n$ such that $\gcd(n,p_r)=1$ for each $r$~\cite{kaidi2021higher}.
Then, one can express the vison by fusion of anyons of each factorized theory, as $v=v_1\times v_2\times\dots\times v_k$. 
For each $v_r$, the twist has the form of $\theta_{v_r}=\exp(\frac{2\pi i s}{p_i^t})$ for some non-negative $s,t\in\mathbb{Z}$. Since $\theta_v=1$, we have $\prod_r \theta_{v_r}=1$, and this implies that $\theta_{v_r}=1$ for each $r$.

By using the vison, the fractional U(1) charge of an anyon $a$ can be expressed as the mutual statistics between $a$ and $v$,
\begin{align}
    e^{2\pi i Q_a}=M_{a,v}.
\end{align}
From this, we see that $Q_{v_r}=0$ mod 1 because $M_{v,v_r}=M_{v_r,v_r}=\theta_{v_r}^2=1$.
Since the bosons $\{v_r\}$ have the trivial mutual statistics with each other, one can condense the anyons $\{v_r\}$. This condensation satisfies the constraint~\eqref{eq:etalagrangian}, and preserves the U(1) symmetry. This can be understood from QFT perspective; the line of $v_r$ generates a 1-form symmetry which is free of mixed 't Hooft anomaly between the U(1) symmetry due to $Q_v=0$, then one can gauge this non-anomalous 1-form symmetry without breaking the U(1) symmetry.

The higher central charge $\xi_n(\mathcal{C}_{p_r})$ with $\gcd(n,p_r)=1$ is invariant under condensation of $v_r$. This can be shown from the fact that $\xi_n$ is given by the phase of the partition function of Reshetikhin-Turaev theory on the lens space $L(n;1)$~\cite{Reshetikhin1991, Turaev+2010, kaidi2021higher} (without background U(1) gauge field),
\begin{align}
    Z_\mathrm{RT}(L(n;1); \mathcal{C}_{p_r})= \frac{1}{\mathcal{D}^2}\sum_{a\in\mathcal{C}_{p_r}}\theta_a^n.
\end{align}
Since the lens space $L(n;1)$ has the trivial background gauge field for the 1-form symmetry generated by the line of $v_r$ when $\gcd(n,p_r)=1$, $Z_\mathrm{RT}(L(n;1); \mathcal{C}_{p_r})$ also gives the partition function of the theory after condensation of $v_r$, which shows the invariance of $\xi_n$ under condensation of $v_r$.

Let us write the theory after condensation of $\{v_r\}$ as
\begin{align}
    \mathcal{C}'=\mathcal{C}'_{p_1}\times \mathcal{C}'_{p_2}\times \dots \times \mathcal{C}'_{p_k}.
\end{align}
Due to the invariance of higher central charge under condensation of $\{v_r\}$, we have $\xi_n(\mathcal{C}'_{p_r})=1$ for all $n$ such that $\gcd(n,p_r)=1$. This implies that each $\mathcal{C}'_{p_r}$ has a gapped boundary, as shown in~\cite{drinfeld2010braided}. After condensing $v_r$, the anyons $a\in\mathcal{C}'_{p_r}$ has trivial U(1) charge $Q_a=0$. So any gapped boundary condition of $\mathcal{C}'_{p_r}$ satisfies the constraint~\eqref{eq:etalagrangian} and preserves U(1) symmetry, since gauging 1-form symmetry free of mixed 't Hooft anomaly does not break U(1) symmetry. Thus, we have shown that $\mathcal{C}'$ admits a gapped boundary, which completes the proof.
\end{proof}

\section{Gapped boundary of (2+1)D fermionic topological phases}
\label{sec:fermionic}
\subsection{Lagrangian algebra anyon for fermionic phases}

We provide a generic theory for topological gapped boundary of (2+1)D topological phases, which are effectively described by spin TQFT. For non-Abelian spin TQFTs, we make a generalization of   Lagrangian algebra anyon for bosonic phases to fermionic cases, and express the gapped boundary condition of spin TQFT in algebraic form. 
In general, (2+1)D spin TQFT is described in terms of a super-modular tensor category $\C$ together with its minimal modular extension $\exC$. The basic description of spin TQFT are reviewed in Appendix~\ref{app:spinTQFT}.

Let us assume that a spin TQFT has a gapped topological boundary condition. For a given 3-manifold, one can cut out a solid torus from it, and equip the resulting 3-manifold with a spin structure. We then introduce a gapped boundary condition on the boundary torus $T^2$. We can think of shrinking the radius of the tube of gapped boundary into a line, then the tube is eventually described by a line operator of spin TQFT. The expression of the line operator depends on the spin structure of the boundary torus $T^2$, and generally described by an object of a minimal modular extension $\Breve{\mathcal{C}}$ of a super-modular tensor category $\C$.

\begin{figure}[htb]
\centering
\includegraphics{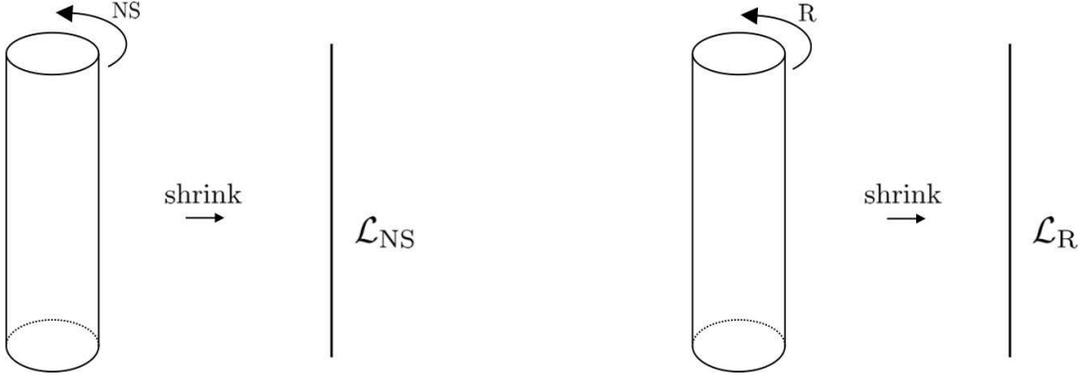}
\caption{One can shrink the tube of gapped boundary into a line, and the resulting object depends on spin structure along the meridian. }
\label{fig:spinshrink}
\end{figure}       

We write spin structure on $T^2$ in the form of $T^2_{\mu,\lambda}$, where $\mu$ is spin structure on the meridian of the tube, and $\lambda$ is that on the longitude.
We also write the minimal modular extension $\Breve{\mathcal{C}}=\Breve{\mathcal{C}}_{\mathrm{NS}}\oplus \Breve{\mathcal{C}}_{\mathrm{R}}$ with $\Breve{\mathcal{C}}_{\mathrm{NS}}=\mathcal{C}$ the NS sector. 

To make an algebraic description of gapped boundary, it is convenient to consider a boundary state $\ket{\L_{\mu,\lambda}}$ on a space $T^2_{\mu,\lambda}$ given by considering a spacetime $T^2_{\mu,\lambda}\times [0,1]$, and then introducing a gapped boundary condition on $T^2_{\mu,\lambda}\times \{1\}$. This defines a state $\ket{\L_{\mu,\lambda}}$ on  $T^2_{\mu,\lambda}\times \{0\}$. 
Equivalently, one can also regard the geometry as a solid torus $D^2\times S^1$, with a thin solid torus cut out at the center of $D^2$. So, the boundary state is thought of as an insertion of a line operator in $D^2\times S^1$ along $S^1$ obtained by a thin tube of gapped boundary.

Based on the description of Hilbert space for spin TQFT explained in Appendix~\ref{app:spinTQFT}, the boundary state with each spin structure $(\mu,\lambda)$ is expressed as follows:
\begin{align}
    \begin{split}
        \ket{\mathcal{L}_{\mathrm{NS},\mathrm{NS}}}&=\sum_{[a]\in\exC_{\mathrm{NS}}/\{1,\psi\}}Z^{\mathrm{NS}}_{0[a]}(\ket{a}+\ket{a\times \psi}) \\
        \ket{\mathcal{L}_{\mathrm{NS},\mathrm{R}}}&=\sum_{[a]\in\exC_{\mathrm{NS}}/\{1,\psi\}}Z^{\mathrm{NS}}_{0[a]}(\ket{a}-\ket{a\times \psi}) \\
     \end{split}
     \label{eq:lagrangianNS}
\end{align}       
\begin{align}
    \begin{split}        
        \ket{\mathcal{L}_{\mathrm{R},\mathrm{NS}}}&=\sum_{[a]\in\exC_{\mathrm{R}}/\{1,\psi\}}Z^{\mathrm{R}}_{0[a]}(\ket{a}+\ket{a\times \psi}) \\
        \ket{\mathcal{L}_{\mathrm{R},\mathrm{R}}}&=\sum_{[a]\in\exC_{\mathrm{R}}/\{1,\psi\}}Z^{\mathrm{R}}_{0[a]}(\ket{a}-\ket{a\times \psi}) \\
    \end{split}
    \label{eq:lagrangianR}
\end{align}
with $Z^{\mathrm{NS}}_{0[a]}, Z^{\mathrm{R}}_{0[a]}$ set to be non-negative integers. 
The reasoning for the expression of the state is that spin structure along a closed curve is measured by inserting a Wilson line of the transparent fermion $\psi$ of $\C$ along the curve. For example, the $\psi$ line along the longitude acts by phase $+1$ (resp.~$-1$) when spin structure along the longitude is NS (resp.~R), hence the state is expressed in the form of $\ket{a}+\ket{a\times \psi}$ (resp.~$\ket{a}-\ket{a\times \psi}$).

For simplicity, we assumed that the R sector $\exC_{\mathrm{R}}$ does not contain an anyon $\sigma$ with fusion rule $\sigma\times \psi = \sigma$ (which is called a ``q-type'' object). For example, when we are interested in spin TQFT with $\U^f$ symmetry which is main interest in this paper, there exists no q-type object in $\exC_{\mathrm{R}}$ since the assignment of U(1) charge on anyons cannot be consistent with the fusion rule $\sigma\times \psi = \sigma$. So, this assumption does not lose generality in that case. The Hilbert space in the presence of q-type objects are illustrated in~\cite{delmastro2021}, and it would be interesting to study boundary conditions of such spin TQFTs.

\subsection{Modular properties of gapped boundary}
\label{sec:STlagrangian}
The boundary states~\eqref{eq:lagrangianNS},~\eqref{eq:lagrangianR} have several important properties under the $S,T$ transformation. Firstly, since the boundary is gapped and topological, the diffeomorphism acting on the gapped boundary at $T^2_{\mu,\lambda}\times\{1\}$ must not change the boundary state $\ket{\mathcal{L}_{\mu,\lambda}}$. Since the torus is equipped with spin structure, the diffeomorphism here means the mapping class group of $T^2_{\mu,\lambda}$ leaving spin structure invariant.

For the case of $\mu=\mathrm{NS}$, the Dehn twist $T$ along the meridian exchanges the spin structure as $T^2_{\mathrm{NS},\mathrm{NS}}\leftrightarrow T^2_{\mathrm{NS},\mathrm{R}}$, so we have $T^2 \ket{\mathcal{L}_{\mathrm{NS},\mathrm{NS}}}=\ket{\mathcal{L}_{\mathrm{NS},\mathrm{NS}}}$. This implies that an anyon that appears in the summand of $\ket{\mathcal{L}_{\mathrm{NS},\mathrm{NS}}}$ must be either a boson or a fermion, since Dehn twist acts on Wilson lines as $\ket{a}\to \theta_a\ket{a}$. 
For convenience, we write $\mathcal{A}_{\mathrm{NS}}=\mathcal{A}_{\mathrm{NS}}^1\sqcup \mathcal{A}_{\mathrm{NS}}^\psi$ as a set of anyons that appear in $\ket{\mathcal{L}_{\mathrm{NS}}}$ (i.e., all $a,a\times\psi\in\exC_{\mathrm{NS}}$ with $Z_{0a}^{\mathrm{NS}}\neq 0$). 
$\mathcal{A}_{\mathrm{NS}}^1\subset \mathcal{A}_{\mathrm{NS}}$ is the set of bosons, and $\mathcal{A}_{\mathrm{NS}}^\psi\subset \mathcal{A}_{\mathrm{NS}}$ is the set of fermions. Note that $ \mathcal{A}_{\mathrm{NS}}^1\cdot \psi=\mathcal{A}_{\mathrm{NS}}^\psi$.

Because of $T\ket{\mathcal{L}_{\mathrm{NS},\mathrm{NS}}}=\ket{\mathcal{L}_{\mathrm{NS},\mathrm{R}}}$, the sum over a class of anyons $[a]=\{a,a\times\psi\}$ in~\eqref{eq:lagrangianNS} is fixed to be the sum over bosons $\mathcal{A}_{\mathrm{NS}}^1$. We then have
\begin{align}
    \begin{split}
        \ket{\mathcal{L}_{\mathrm{NS},\mathrm{NS}}}=\sum_{a\in\mathcal{A}_{\mathrm{NS}}^1}Z^{\mathrm{NS}}_{0a}(\ket{a}+\ket{a\times \psi}) \\
        \ket{\mathcal{L}_{\mathrm{NS},\mathrm{R}}}=\sum_{a\in\mathcal{A}_{\mathrm{NS}}^1}Z^{\mathrm{NS}}_{0a}(\ket{a}-\ket{a\times \psi}) \\
     \end{split}
\end{align}
where we define $Z_{0a}^{\mathrm{NS}}=Z_{0,a\times\psi}^{\mathrm{NS}}=Z_{0[a]}^{\mathrm{NS}}$.

For the case of $\mu=\mathrm{R}$, the Dehn twist along the meridian leaves the spin structure invariant, so we have $T \ket{\mathcal{L}_{\mathrm{R},*}}=\ket{\mathcal{L}_{\mathrm{R},*}}$. This means that an anyon that appears in the summand of $\ket{\mathcal{L}_{\mathrm{R},*}}$ must be a boson. For convenience, we again write $\mathcal{A}_{\mathrm{R}}=\mathcal{A}_{\mathrm{R}}^e\sqcup \mathcal{A}_{\mathrm{R}}^m$ as a set of anyons that appear in $\ket{\mathcal{L}_{\mathrm{R}}}$ (i.e., all $a,a\times\psi\in\exC_{\mathrm{R}}$ with $Z_{0a}^{\mathrm{R}}\neq 0$). The decomposition $\mathcal{A}_{\mathrm{R}}^e\sqcup \mathcal{A}_{\mathrm{R}}^m$ is taken to satisfy
$\mathcal{A}_{\mathrm{R}}^e\times \psi=\mathcal{A}_{\mathrm{R}}^m$.
We note that there is no canonical way to perform the decomposition into $\mathcal{A}_{\mathrm{R}}^e$ and $\mathcal{A}_{\mathrm{R}}^m$ unlike the case of NS sector, since all anyons of $\mathcal{A}_{\mathrm{R}}$ are bosons.
We have
\begin{align}
    \begin{split}        
        \ket{\mathcal{L}_{\mathrm{R},\mathrm{NS}}}&=\sum_{a\in\mathcal{A}_{\mathrm{R}}^e}Z^{\mathrm{R}}_{0a}(\ket{a}+\ket{a\times \psi}) \\
        \ket{\mathcal{L}_{\mathrm{R},\mathrm{R}}}&=\sum_{a\in\mathcal{A}_{\mathrm{R}}^e}Z^{\mathrm{R}}_{0a}(\ket{a}-\ket{a\times \psi}) \\
    \end{split}
    \label{eq:Rstates}
\end{align}
 where we define $Z_{0a}^{\mathrm{R}}=Z_{0,a\times\psi}^{\mathrm{R}}=Z_{0[a]}^{\mathrm{R}}$.

\begin{figure}[htb]
\centering
\includegraphics{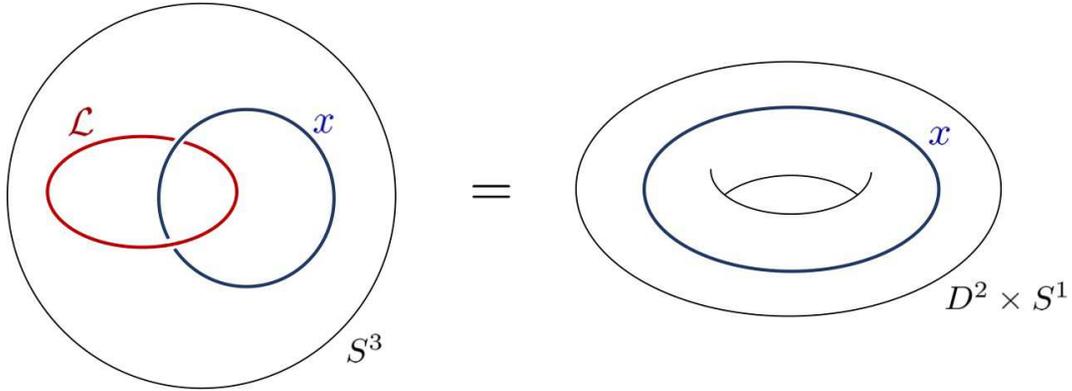}
\caption{Hopf link between the tube $\mathcal{L}$ of a gapped boundary and an anyon, which is topologically equivalent to the partition function on $D^2\times S^1$.}
\label{fig:hopflink}
\end{figure}

Next, we discuss the property of the boundary states under the $S$ transformation. In the spacetime manifold $S^3$, let us consider a Hopf link between the thin tube of gapped boundary and a line operator $x$, as described in Fig.~\ref{fig:hopflink}. This Hopf link can be evaluated in two ways. On one hand, the amplitude of the Hopf link is computed by the modular $S$-matrix of a modular category $\exC$ between two states on $T^2$, the boundary state $\ket{\mathcal{L}}$ and $\ket{x}$ respectively. 
On the other hand, one can enlarge the radius of the boundary tube, which makes the geometry $D^2\times S^1$ with the gapped boundary on $T^2$. It then reduces to a partition function on $D^2\times S^1$ with a $x$ line inserted along $S^1$. Comparing these two distinct expressions put a nontrivial constraint on the vectors $Z_{0a}$. 

The analysis of $S$-matrix action on boundary states requires a lengthy discussion done by cases for spin structure of the spatial torus, and the detail is relegated to Appendix~\ref{app:smatrix}. Here we summarize the $S$-matrix actions on $Z_{0a}$ obtained as follows:
\begin{align}
\begin{split}
    \sum_{b\in\mathcal{A}_{\mathrm{NS}}^1}2S_{ab}Z_{0b}^{\mathrm{NS}}&=Z_{0a}^{\mathrm{NS}} \quad \text{for $a\in\exC_{\mathrm{NS}}$,} \qquad
    \sum_{b\in\mathcal{A}_{\mathrm{NS}}^\psi}2S_{ab}Z_{0b}^{\mathrm{NS}}=Z_{0a}^{\mathrm{NS}} \quad \text{for $a\in\exC_{\mathrm{NS}}$} 
    \end{split}
    \label{eq:SforNSNS}
\end{align}

\begin{align}
\begin{split}
        \sum_{b\in\mathcal{A}_{\mathrm{NS}}^1}2S_{ab}Z_{0b}^{\mathrm{NS}}&=Z_{0a}^{\mathrm{R}} \quad \text{for $a\in\exC_{\mathrm{R}}$,} \qquad
        \sum_{b\in\mathcal{A}_{\mathrm{NS}}^\psi}2S_{ab}Z_{0b}^{\mathrm{NS}}=-Z_{0a}^{\mathrm{R}} \quad \text{for $a\in\exC_{\mathrm{R}}$}
        \end{split}
            \label{eq:SforNSR}
    \end{align}

    \begin{align}
    \begin{split}
        \sum_{b\in\mathcal{A}_{\mathrm{R}}^e}2S_{ab}Z_{0b}^{\mathrm{R}}&=Z_{0a}^{\mathrm{NS}} \quad \text{if $a\in \mathcal{A}_{\mathrm{NS}}^1$}, \qquad
        \sum_{b\in\mathcal{A}_{\mathrm{R}}^e}2S_{ab}Z_{0b}^{\mathrm{R}}=-Z_{0a}^{\mathrm{NS}} \quad \text{if $a\in \mathcal{A}_{\mathrm{NS}}^\psi$} \\
        \sum_{b\in\mathcal{A}_{\mathrm{R}}^m}2S_{ab}Z_{0b}^{\mathrm{R}}&=Z_{0a}^{\mathrm{NS}} \quad \text{if $a\in \mathcal{A}_{\mathrm{NS}}^1$} \qquad
        \sum_{b\in\mathcal{A}_{\mathrm{R}}^m}2S_{ab}Z_{0b}^{\mathrm{R}}=-Z_{0a}^{\mathrm{NS}} \quad \text{if $a\in \mathcal{A}_{\mathrm{NS}}^\psi$} \\
        \end{split}
        \label{eq:SforRNS}
    \end{align}
   \eqref{eq:SforRNS} also holds when $a$ is not an element of $\mathcal{A}_{\NS}$, where both lhs and rhs becomes zero.
    
        \begin{align}
    \begin{split}
        \sum_{b\in\mathcal{A}_{\mathrm{R}}^e}2S_{ab}Z_{0b}^{\mathrm{R}}&=Z_{0a}^{\mathrm{R}} \quad \text{if $a\in \mathcal{A}_{\mathrm{R}}^e$,} \qquad 
        \sum_{b\in\mathcal{A}_{\mathrm{R}}^e}2S_{ab}Z_{0b}^{\mathrm{R}}=-Z_{0a}^{\mathrm{R}} \quad \text{if $a\in \mathcal{A}_{\mathrm{R}}^m$} \\
        \sum_{b\in\mathcal{A}_{\mathrm{R}}^m}2S_{ab}Z_{0b}^{\mathrm{R}}&=-Z_{0a}^{\mathrm{R}} \quad \text{if $a\in \mathcal{A}_{\mathrm{R}}^e$,}\qquad 
        \sum_{b\in\mathcal{A}_{\mathrm{R}}^m}2S_{ab}Z_{0b}^{\mathrm{R}}=Z_{0a}^{\mathrm{R}} \quad \text{if $a\in \mathcal{A}_{\mathrm{R}}^m$} \\
        \end{split}
         \label{eq:SforRR}
    \end{align}
    \eqref{eq:SforRR} also holds when $a$ is not an element of $\mathcal{A}_{\R}$, where both lhs and rhs becomes zero.

\subsection{Stacking with toric code $D(\Z_2)$}
\label{sec:DZ2}
In the above study of the $S$-matrix action on boundary states, we obtained a bunch of equations in the form of $2SZ=\pm Z$ for each spin structure. We can express the above many equations in a much simpler way, with help of toric code $D(\Z_2)$ with anyons $\{1,\psi, e, m\}$ and $S$-matrix $S^{\mathrm{tc}}$. The above  properties of the boundary states under the action of $S$-matrix~\eqref{eq:SforNSNS}~\eqref{eq:SforNSR}~\eqref{eq:SforRNS}~\eqref{eq:SforRR} are then neatly summarized as
\begin{align}
    \begin{split}
        \sum_{b\in\mathcal{A}_{\mathrm{NS}}^x}4S_{ab}(S^{\mathrm{tc}})_{xy}Z_{0b}^{\mathrm{NS}}&=Z_{0a}^{\mathrm{NS}} \quad \text{for $a\in\mathcal{A}_{\mathrm{NS}}^y$} \\
        \sum_{b\in\mathcal{A}_{\mathrm{NS}}^x}4S_{ab}(S^{\mathrm{tc}})_{xy}Z_{0b}^{\mathrm{NS}}&=Z_{0a}^{\mathrm{R}} \quad \text{for $a\in\mathcal{A}_{\mathrm{R}}^y$} \\
        \sum_{b\in\mathcal{A}_{\mathrm{R}}^x}4S_{ab}(S^{\mathrm{tc}})_{xy}Z_{0b}^{\mathrm{R}}&=Z_{0a}^{\mathrm{NS}} \quad \text{for $a\in\mathcal{A}_{\mathrm{NS}}^y$} \\
        \sum_{b\in\mathcal{A}_{\mathrm{R}}^x}4S_{ab}(S^{\mathrm{tc}})_{xy}Z_{0b}^{\mathrm{R}}&=Z_{0a}^{\mathrm{R}} \quad \text{for $a\in\mathcal{A}_{\mathrm{R}}^y$} \\
    \end{split}
\end{align}
with $x,y\in D(\Z_2)$. 
Then, let us define a new set of anyons $\widetilde{\mathcal{A}}_{\mathrm{NS}}\subset \exC_{\mathrm{NS}}\boxtimes 
D(\Z_2)$ as 
\begin{align}
   \widetilde{\mathcal{A}}_{\mathrm{NS}}:= \{(a,1)|a\in{\mathcal{A}}_{\mathrm{NS}}^1\} \sqcup\{(a,\psi)|a\in{\mathcal{A}}_{\mathrm{NS}}^\psi\} 
\end{align}
Here $\boxtimes$ denotes the Deligne product, which physically means stacking two independent theories.

 We also define $\widetilde{\mathcal{A}}_{\mathrm{R}}\subset \exC_{\mathrm{R}}\boxtimes D(\Z_2)$ as
\begin{align}
  \widetilde{\mathcal{A}}_{\mathrm{R}}:= \{(a,e)|a\in{\mathcal{A}}_{\mathrm{R}}^e\} \sqcup\{(a,m)|a\in{\mathcal{A}}_{\mathrm{R}}^m\}
\end{align}
We can then write the equations in terms of $S$-matrix of $\exC\boxtimes 
D(\Z_2)$ as
\begin{align}
    \begin{split}
        \sum_{b\in\widetilde{\mathcal{A}}_{\mathrm{NS}}}4(S^{\exC\boxtimes D(\Z_2)})_{ab}Z_{0b}^{\mathrm{NS}}&=2Z_{0a}^{\mathrm{NS}} \quad \text{for $a\in\tilde{\mathcal{A}}_{\mathrm{NS}}$} \\
        \sum_{b\in\widetilde{\mathcal{A}}_{\mathrm{NS}}}4(S^{\exC\boxtimes D(\Z_2)})_{ab}Z_{0b}^{\mathrm{NS}}&=2Z_{0a}^{\mathrm{R}} \quad \text{for $a\in\tilde{\mathcal{A}}_{\mathrm{R}}$} \\
        \sum_{b\in\widetilde{\mathcal{A}}_{\mathrm{R}}}4(S^{\exC\boxtimes D(\Z_2)})_{ab}Z_{0b}^{\mathrm{R}}&=2Z_{0a}^{\mathrm{NS}} \quad \text{for $a\in\tilde{\mathcal{A}}_{\mathrm{NS}}$} \\
        \sum_{b\in\widetilde{\mathcal{A}}_{\mathrm{R}}}4(S^{\exC\boxtimes D(\Z_2)})_{ab}Z_{0b}^{\mathrm{R}}&=2Z_{0a}^{\mathrm{R}} \quad \text{for $a\in\tilde{\mathcal{A}}_{\mathrm{R}}$} \\
    \end{split}
\end{align}
Finally, by writing the combined vector $\widetilde{Z}:=Z_{\mathrm{NS}}\oplus Z_{\mathrm{R}}$ and $\widetilde{\mathcal{A}}:=\widetilde{\mathcal{A}}_{\mathrm{NS}}\sqcup \widetilde{\mathcal{A}}_{\mathrm{R}}$, we obtain
\begin{align}
    \sum_{b\in\widetilde{\mathcal{A}}}(S^{\exC\boxtimes D(\Z_2)})_{ab}\widetilde{Z}_{0b}&=\widetilde{Z}_{0a} \quad \text{for $a\in\widetilde{\mathcal{A}}$}
\end{align}
and it is not hard to see that the lhs becomes zero when $a$ is not an element of $\widetilde{\mathcal{A}}$.
That is, the vector $\widetilde{Z}$ gives an eigenvector of the $S$-matrix of $\exC\boxtimes D(\Z_2)$, $S\widetilde{Z}=\widetilde{Z}$. Since the anyons in $\widetilde{\mathcal{A}}$ are all bosons, we also have $T\widetilde{Z}=\widetilde{Z}$. So, a newly constructed object
$\widetilde{\mathcal{L}}:=\sum_{a\in\widetilde{\mathcal{A}}}\widetilde{Z}_{0a}a$
 satisfies the crucial property for the Lagrangian algebra anyon of a bosonic TQFT $\exC\boxtimes D(\Z_2)$.
Hence, we expect that $\widetilde{Z}$ describes a bosonic gapped boundary of a bosonic modular theory $\exC\boxtimes D(\Z_2)$. 

Physically, the gapped boundary of a fermionic theory $\mathcal{C}$ is regarded as a gapped interface between $\mathcal{C}$ and a trivial fermionic invertible phase. By gauging fermion parity $\Z_2^f$ of the whole system in the presence of the gapped boundary, we obtain a bosonic gapped interface between $\Breve{\mathcal{C}}$ and a $\Z_2$ toric code $D(\mathbb{Z}_2)$, where the toric code $D(\mathbb{Z}_2)$ corresponds to gauging a trivial fermionic invertible phase. Since $D(\mathbb{Z}_2)=\overline{D(\mathbb{Z}_2)}$, it implies the existence of the gapped boundary for a bosonic theory $\exC\boxtimes D(\Z_2)$. In the next subsection, we confirm this physical intuition by checking the other additional properties of Lagrangian algebra anyon.  

\subsection{Additional properties of Lagrangian algebra anyon}
\begin{figure}[htb]
\centering
\includegraphics{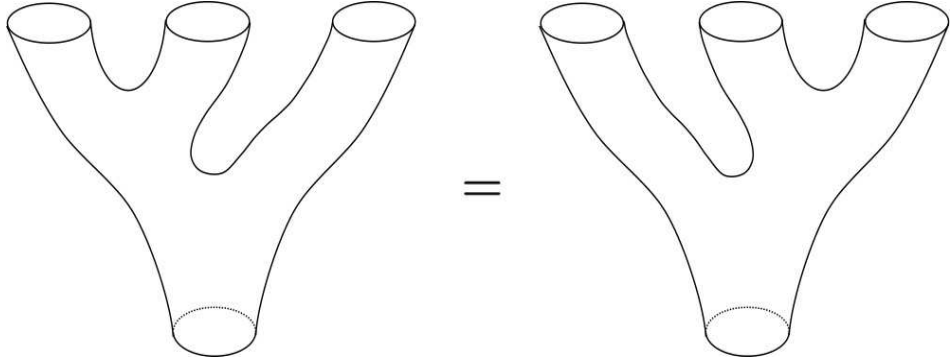}
\caption{We carve out the junction of tubes, and then introduce gapped boundary. The $F$-move of tubes is obviously trivial since it can be realized by diffeomorphism. }
\label{fig:spinFmove}
\end{figure}

Here we study the additional constraint of the anyon $\widetilde{\mathcal{L}}=\sum_{a\in\widetilde{\mathcal{A}}}\widetilde{Z}_{0a}a$ satisfied by Lagrangian algebra anyon represented in~\eqref{eq:lagrangianFboson},~\eqref{eq:lagrangianRboson}. The first constraint involves $F$-move of anyons, which is obtained by cutting out the junction of tubes from the spacetime and introducing the gapped boundary, as shown in Fig.~\ref{fig:spinFmove}.
Note that each tube of gapped boundary is equipped with spin structure along its meridian, 
and write the corresponding object as 
\begin{align}
    \mathcal{L}_{\NS}= \bigoplus_{a\in\mathcal{A}_{\NS}}Z_{0a}^{\NS}a, \quad  \mathcal{L}_{\R}= \bigoplus_{a\in\mathcal{A}_{\R}}Z_{0a}^{\R}a
\end{align}

Since the $F$-move can be realized by topological deformation of gapped boundary, $F$-move acts trivially on the tube of gapped boundary. So, writing $\ket{\mu^{pq}_{r}}\in V^{\mathcal{L}_p\mathcal{L}_q}_{\mathcal{L}_r}$ as a fusion vector of three tubes of gapped boundaries, we have
\begin{align}
    (F^{\mathcal{L}_{p}\mathcal{L}_{q}\mathcal{L}_{r}}_{\mathcal{L}_{u}})_{\mathcal{L}_{s}, \mathcal{L}_{t}}\cdot \ket{\mu^{pq}_{s}}\otimes  \ket{\mu^{sr}_{u}} = \ket{\mu^{pt}_{u}}\otimes  \ket{\mu^{qr}_{t}}
\end{align}
Since the $F$-matrices of $D(\Z_2)$ is also completely trivial, for a fusion vector $\ket{\widetilde{\mu}}\in V^{\widetilde{\mathcal{L}}\widetilde{\mathcal{L}}}_{\widetilde{\mathcal{L}}}$ we immediately have~\eqref{eq:lagrangianFboson}
\begin{align}
    (F^{\widetilde{\mathcal{L}}\widetilde{\mathcal{L}}\widetilde{\mathcal{L}}}_{\widetilde{\mathcal{L}}})_{\widetilde{\mathcal{L}}, \widetilde{\mathcal{L}}}\cdot \ket{\widetilde{\mu}}\otimes  \ket{\widetilde{\mu}} = \ket{\widetilde{\mu}}\otimes \ket{\widetilde{\mu}}
    \label{eq:lagrangianF}
\end{align}

\begin{figure}[htb]
\centering
\includegraphics{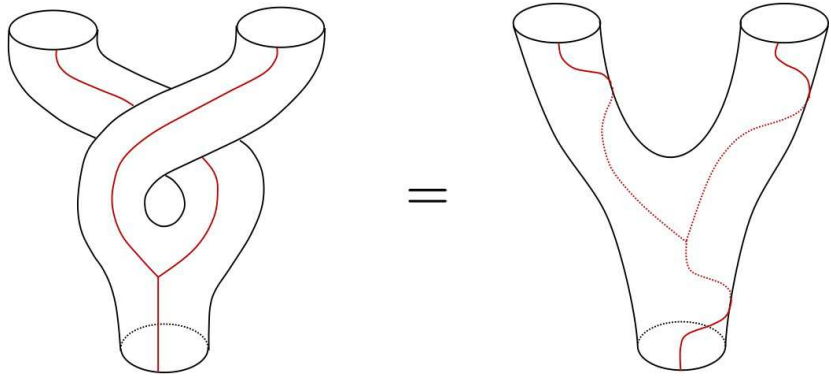}
\caption{$R$-move for a junction of tubes. One can realize the $R$-move by a diffeomorphism up to a ``half-twist'' on each tube, represented by $\pi$ rotation of a red line along its meridian.}
\label{fig:spinRmove}
\end{figure}

The next constraint has to do with $R$-move of anyons, which is a bit more involved. This constraint is obtained by considering a ``half-twist'' of the junction of gapped boundaries as shown in Fig.~\ref{fig:spinRmove}. 
This process amounts to twisting each tube of gapped boundary by $\pi$ along the meridian. We write the $\pi$ twist as $T^{\frac{1}{2}}$ since performing it twice gives a single Dehn twist $T$. $T^{\frac{1}{2}}$ acts diagonally on anyons, and has the form of
\begin{align}
\begin{split}
    (T^{\frac{1}{2}})_{a,a}&= 1 \quad \text{for $a\in\mathcal{A}_{\NS}^{1}, \mathcal{A}_{\R}^{e}, \mathcal{A}_{\R}^{m}$} \\
    (T^{\frac{1}{2}})_{a,a}&= i \quad \text{for $a\in\mathcal{A}_{\NS}^{\psi}$} \\
    \end{split}
\end{align}

By writing the objects for gapped boundary as $\mathcal{L}_{\NS}=\mathcal{L}_{\NS}^1+\mathcal{L}_{\NS}^\psi$ and $\mathcal{L}_{\R}=\mathcal{L}_{\R}^e+\mathcal{L}_{\R}^m$ with
\begin{align}
    \mathcal{L}^1_{\NS}= \bigoplus_{a\in\mathcal{A}^1_{\NS}}Z_{0a}^{\NS}a, \quad  \mathcal{L}^{\psi}_{\NS}= \bigoplus_{a\in\mathcal{A}_{\NS}^{\psi}}Z_{0a}^{\NS}a
\end{align}
\begin{align}
    \mathcal{L}^e_{\R}= \bigoplus_{a\in\mathcal{A}^e_{\R}}Z_{0a}^{\R}a, \quad  \mathcal{L}^{m}_{\R}= \bigoplus_{a\in\mathcal{A}_{\R}^{m}}Z_{0a}^{\R}a
\end{align}
the action of $R$-move on the fusion vector $\ket{\mu^{xy}_{z}}$ for $x,y,z\in D(\Z_2)$ is then expressed as
\begin{align}
    R^{\mathcal{L}^{x}\mathcal{L}^{y}}_{\mathcal{L}^{z}}\ket{\mu^{xy}_{z}}=(T^{-\frac{1}{2}})_{x,x}(T^{-\frac{1}{2}})_{y,y}(T^{\frac{1}{2}})_{z,z}\ket{\mu^{xy}_{z}}
\end{align}
One can see that this gives the same $R$ matrix as that of $\overline{D(\Z_2)}$, $R^{\mathcal{L}^{x}\mathcal{L}^{y}}_{\mathcal{L}^{z}}=(R^{xy}_{z})^*$ by picking a proper $R^{xy}_{z}$ of $D(\Z_2)$ by suitable vertex basis transformation~\eqref{eq:vertexBasisTransformation}. Hence, one can see that the $R$-matrix element of $\widetilde{\mathcal{L}}$ for $\exC\boxtimes D(\Z_2)$ is trivial and~\eqref{eq:lagrangianRboson} follows,
\begin{align}
    R^{\widetilde{\mathcal{L}}^{x}\widetilde{\mathcal{L}}^{y}}_{\widetilde{\mathcal{L}}^{z}}\ket{\widetilde{\mu}^{xy}_{z}}=R^{\mathcal{L}^{x}\mathcal{L}^{y}}_{\mathcal{L}^{z}}R^{xy}_{z}\ket{\widetilde{\mu}^{xy}_{z}}=\ket{\widetilde{\mu}^{xy}_{z}}
\end{align}

\subsection{U(1)$^f$ symmetry-preserving gapped boundary}
\label{sec:U1f}
In this section, we study symmetry-preserving gapped boundary of (2+1)D spin TQFT.  For simplicity, we focus on global symmetry $G_f=\U^f$ given by the symmetry extension
    \begin{align}
        \Z_2^f\to G_f\to G_b,
    \end{align}    
    where $G_b=\U$ is the bosonic symmetry group. Note that $G_b=\mathbb{R}/(\Z/2)$ has periodicity 1/2, since we want the fermionic group $G_f=\mathbb{R}/\Z$ to have periodicity 1.
    
    As reviewed in Appendix~\ref{fermSymFracSec}, in order to describe a spin TQFT with $G_f$ symmetry, we first take $G_b$ symmetry on the bosonic theory given by a minimal modular extension $\exC = \exC_{\NS}\oplus \exC_{\R}$ as an input. 
    In general, a spin TQFT is constructed by performing a process of ``fermion condensation'' for a given bosonic TQFT $\exC$, see Appendix~\ref{app:spinTQFT} for a review of spin TQFT.
    This process corresponds to gauging 1-form symmetry generated by a Wilson line of a transparent fermion $\psi$, and the resulting theory has a dual $\Z_2^f$ fermion parity symmetry. So, after fermion condensation starting with the bosonic theory $\exC$, we obtain a spin TQFT with $G_f$ symmetry.
    
    We demand that~\cite{bulmashSymmFrac,aasen21ferm} (see Appendix~\ref{fermSymFracSec} for an explanation)
        \begin{align}
            U_{\bf g}(\psi,\psi;1) &= 1\\
            \eta_{\psi}({\bf g}, {\bf h}) &= \omega_2({\bf g}, {\bf h})
            \label{eq:psifrac}
        \end{align}
        where $[\omega_2] \in H^2(BG_b,\mathbb{Z}_2)$ is the cohomology class specifying $G_f$ as a group extension of $G_b$ by $\mathbb{Z}_2$.

The $G_b$ action of $\exC$ is realized as a topological symmetry of BTC which preserves the $\Z_2$ grading $\exC=\exC_\NS\oplus\exC_\R$. So, it induces a topological symmetry of the NS sector $\exC_\NS=\C$. 
We can define fractional charges of anyons in $\C$ analogously to the bosonic case. For a fixed anyon $a$, let $n$ be the smallest integer such that $a^n$ contains the identity as a fusion product. Choose a sequence of anyons $a, a^2,\dots a^n=1$ such that $a\times a^k$ contains $a^{k+1}$ as a fusion product. Then define a fractional charge $Q_a\in\mathbb{R}/(2\Z)$ as \cite{bulmashSymmFrac}
\begin{align}
    e^{\pi i Q_a}:= \prod_{m=1}^{n-1}\eta_a\left(\frac{1}{2n}, \frac{m}{2n}\right)U_{\frac{1}{2n}}(a,a^m; a^{m+1}),
    \label{eq:chargedefspinc}
\end{align}
where the elements of $G_b$ is labeled by numbers in $[0,1/2)$. Since $[\rho]$ is the identity map, there is a gauge in which we can set $U = 1$, so that 
    $\eta$ satisfies ${\eta_c({\bf g}, {\bf h})}=\eta_a({\bf g}, {\bf h})\eta_b({\bf g}, {\bf h})$ when $N^{c}_{ab}>0$. Then one can see that $e^{\pi i Q_a}$ satisfies
      \begin{align}
    e^{\pi i Q_a}e^{\pi i Q_b}=e^{\pi i Q_c}\quad \text{when $N_{ab}^c\neq 0$}.
    \label{eq:fusionchargespinc}
\end{align}  
Since $\omega_2({\bf g}, {\bf h})=e^{2\pi i (\mathbf{g}+\mathbf{h}-[\mathbf{g}+\mathbf{h}]_{1/2})}$, the fractional charge of a transparent fermion $\psi$ satisfies
\begin{align}
    Q_{\psi}=1 \mod 2.
    \label{eq:Qpsi1}
\end{align}
Since $\U$ symmetry does not permute anyons we can set the gauge $U=1$, then we can write the phases $\eta_a({\bf g}, {\bf h})$ as
\begin{align}
    \eta_a({\bf g},{\bf h}) = M_{a, \mathfrak{t}({\bf g},{\bf h})} ,
\end{align}
for an Abelian anyon $\mathfrak{t}({\bf g},{\bf h})\in Z^2(B\U, \mathcal{B})$ where $\mathcal{B}$ is the set of Abelian anyons in $\exC$. A representative 2-cocycle $\mathfrak{t}$ is given by $\mathfrak{t}({\bf g}, {\bf h}) = v^{2({\bf g} + {\bf h} - [ {\bf g} + {\bf h}]_{1/2})} ,$
where $v \in \mathcal{B}$ is referred to as the vison, $\mathbf{g},\mathbf{h}\in \mathbb{R}/(\Z/2)$ takes the values in $[0,1/2)$, and $[\mathbf{g}+\mathbf{h}]_{1/2}$ means the sum mod 1/2. In particular, since we have $M_{\psi,v}=-1$ in order to satisfy~\eqref{eq:psifrac}, the vison must be in the R sector $v\in\exC_{\R}$. Using the vison, the fractional charge~\eqref{eq:chargedefspinc} can be expressed as
\begin{align}
    e^{\pi i Q_a}=M_{a,v}.
    \label{eq:chargevisonspinc}
\end{align}

Now we are prepared to discuss the symmetry-preserving gapped boundary of spin TQFT with U(1)$^f$ symmetry. Analogously to the bosonic case in Sec.~\ref{sec:symmetry}, we expect that the partition function of the bulk-boundary system without operator insertions is invariant under background gauge transformations, since it is expected that both the bulk and boundary cannot have 't Hooft anomaly in order to admit a symmetry-preserving gapped boundary~\cite{Thorngren2021end}.
Though we are not aware of a rigorous argument for this in fermionic case due to the presence of subtle anomaly beyond group cohomology intrinsic in fermionic systems~\cite{gu2014, Kapustin:2014dxa, qingrui, Thorngren2018bosonization, BCHM2021classification}, here we assume that the partition function is invariant under background $G_f$ gauge transformations.

\begin{figure}[htb]
\centering
\includegraphics{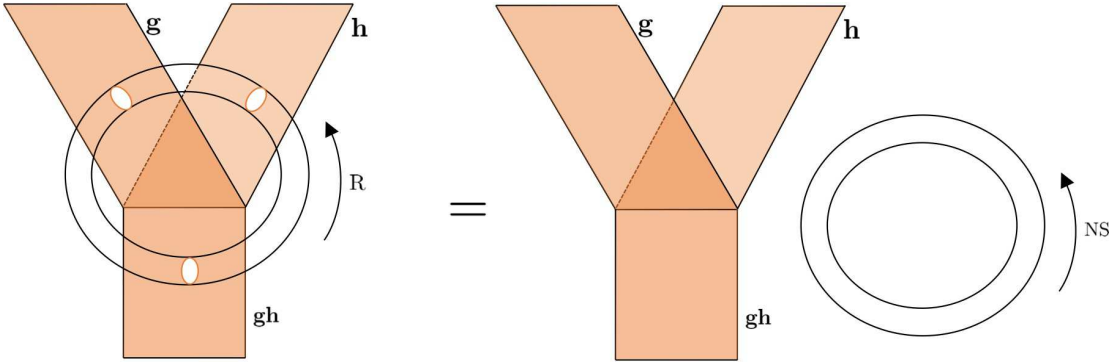}
\caption{Insertion of a tube of gapped boundary through symmetry defects with a junction. By background gauge transformation, one can move the junction away from the tube, which shifts spin structure of the tube along its longitude.}
\label{fig:spintubedefect}
\end{figure}

Then, similar to what we have done for bosonic case in Sec.~\ref{sec:symmetry}, we think of carving out a solid torus from a spacetime 3-manifold, and introduce a gapped boundary condition on the boundary of the solid torus.  We consider a junction of symmetry defects $\mathbf{g}, \mathbf{h}, \mathbf{g}+\mathbf{h}$ piercing the solid torus, see Fig.~\ref{fig:spintubedefect}
When the symmetry defect ends on boundary, it realizes a symmetry defect of a gapped boundary theory. 

By performing background gauge transformations, one can move the symmetry defects away from the tube of gapped boundary without producing a phase. By comparing the configuration before and after the gauge transformation, we get a constraint on the symmetry fractionalization data of the Lagrangian algebra anyon. In particular, consider a flat U(1)$^f$ background gauge field realized by a junction of three defects, where $\mathbf{g}+\mathbf{h}-[\mathbf{g}+\mathbf{h}]=1/2$. In that case, due to the symmetry extension $\Z_2^f\to \U^f\to \U$, the junction bounds a vortex for fermion parity $\Z_2^f$ symmetry, hence the tube of gapped boundary before and after gauge transformation carry a different spin structure along the longitude. That is, suppose that the initial tube has  $(\mu,\lambda)$=(NS,R), then the final tube has (NS,NS). 

By shrinking the tube of gapped boundary, it follows that for the Lagrangian algebra anyon in NS sector $\mathcal{L}_{\NS}^1+\mathcal{L}_{\NS}^\psi$, moving the tube across the junction of defects has the effect of acting on $\mathcal{L}_{\NS}^\psi$ by $-1$ phase, while acting on $\mathcal{L}_{\NS}^1$ trivially. Therefore, we conclude that $\eta_a(\mathbf{g},\mathbf{h})$ has the form of
\begin{align}
    \eta_{\mathcal{L}_{\NS}^1}(\mathbf{g},\mathbf{h})=1, \quad \eta_{\mathcal{L}_{\NS}^\psi}(\mathbf{g},\mathbf{h})=\omega_2(\mathbf{g},\mathbf{h}).
\end{align}
In the gauge where $U=1$, it implies that the fractional charge is trivial for anyons in $\mathcal{A}_{\NS}^1$,
\begin{align}
    e^{\pi i Q_a}=M_{a,v}=1 \quad \text{when $a\in \mathcal{A}_{\NS}^1$}.
\end{align}
We can then show that the vison is condensable $v\in \mathcal{A}_{\R}$, by using the modular property of the Lagrangian algebra anyon~\eqref{eq:SforNSR} and $S_{v,b} = \frac{d_b M^*_{v,b}}{\mathcal{D}}$,
\begin{align}
    Z_{0v}^{\mathrm{R}}=\sum_{b\in\mathcal{A}_{\mathrm{NS}}^1}2S_{v,b}Z_{0b}^{\mathrm{NS}}=\frac{1}{\mathcal{D}}\sum_{b\in\mathcal{A}_{\mathrm{NS}}^1}2d_bZ_{0b}^{\mathrm{NS}} = Z_{00}^{\NS} > 0.
\end{align}
Here, $Z_{00}^{\NS}> 0$ follows from the observation that the gapped boundary on a $S^2$ is topologically deformed to a thin tube with NS spin structure ending on the north and south pole, which means that $\mathrm{Hom}(\mathcal{L}_{\NS},1)$ is non-vanishing. Since the anyons in $\mathcal{A}_{\R}$ are all bosons as shown in Sec.~\ref{sec:STlagrangian}, the vison must be a boson, $\theta_v=1$.

\section{Higher version of $c_-$ and Hall conductivity}
\label{sec:higher}
In this section, we focus on fermionic topological phases with U(1)$^f$ symmetry. We prove the following statement:
\begin{theorem}
Fermionic topological phase with U(1)$^f$ symmetry described by a super-modular category $\mathcal{C}$ has a U(1)$^f$ symmetric gapped boundary only if the quantity $\zeta_n$ defined as
\begin{align}
    \zeta_n:=\frac{\sum_{a\in\mathcal{C}}e^{i\pi Q_a} d_a^2\theta_a^n}{|\sum_{a\in\mathcal{C}}e^{i\pi Q_a} d_a^2\theta_a^n|}
\end{align}
becomes 1 for all $n$ such that $\gcd(n,N_{\mathrm{FS}})=1$.
\end{theorem}
Here, $N_{\mathrm{FS}}$ is the Frobenius-Schur exponent of super-modular tensor category $\mathcal{C}$.

\begin{proof}
For a given super-modular category $\C$, we pick a minimal modular extension $\exC$ where symmetry fractionalization data of $\C$ lifts to $\U$ symmetry fractionalization of $\exC$, which is guaranteed to exist as discussed in Sec.~\ref{sec:U1f}.

Let $\mathcal{B}$ be the group of Abelian anyons in $\C$.
We firstly show that for any $a\in\mathcal{B}$, $N_{\mathrm{FS}}$ is integer multiple of $N_a$, where we define $N_a$ as the smallest integer that $a^{N_a}=1$.
We prove this statement by contradiction. Suppose that there exists an Abelian anyon $a\in\mathcal{B}$ such that $N_a$ does not divide $N_{\mathrm{FS}}$.
Note that $\theta_{a\times x}=\theta_a\theta_x M_{a,x}$ for any $x\in\mathcal{C}$, which means that $M_{a,x}$ is $N_{\mathrm{FS}}$-th root of unity. At the same time, $M_{a,x}$ is $N_{a}$-th root of unity since $M_{a,x}^{N_a}=M_{a^{N_a},x}=1$, so $M_{a,x}^{\gcd(N_{\mathrm{FS}},N_a)}=M_{a^{\gcd(N_{\mathrm{FS}},N_a)},x}=1$ for all $x\in\mathcal{C}$. 

Since $N_a$ does not divide $N_{\mathrm{FS}}$, $\gcd(N_{\mathrm{FS}},N_a)< N_a$ and $a^{\gcd(N_{\mathrm{FS}},N_a)}\neq 1$. Hence, due to super-modularity, we must have $a^{\gcd(N_{\mathrm{FS}},N_a)}=\psi$. This means that $\gcd(N_{\mathrm{FS}},N_a)=N_a/2$, so $N_{\mathrm{FS}}$ is odd multiple of $N_a/2$, and hence $\gcd(N_{\mathrm{FS}},2N_a)=N_a/2$. Since $\theta_a^{2N_a}=M_{a,a}^{N_a}=1$, $\theta_a$ is $\gcd(N_{\mathrm{FS}},2N_a)=N_a/2$-th root of unity, so $\theta_{a^{N_a/2}}=(\theta_a)^{(\frac{N_a}{2})^2}=1$. This contradicts with $a^{\gcd(N_{\mathrm{FS}},N_a)}=\psi$, so $N_{\mathrm{FS}}$ must be an integer multiple of $N_a$.

For a vison $v\in \exC_{\R}$ in~\eqref{eq:chargevisonspinc}, let $N_v$ be the smallest integer satisfying $v^{N_v}=1$. 
Since $v^2\in\mathcal{B}$, $N_v$ is expressed as $N_v=2N_{v^2}$ where $N_{v^2}$ divides $N_{\mathrm{FS}}$. When $n$ satisfies $\gcd(n,N_{\mathrm{FS}})=1$, we also have $\gcd(n, N_v)=1$, because $n$ must be odd since $N_{\mathrm{FS}}$ is even due to the presence of the fermion $\psi$, and $\gcd(n,N_{v^2})=1$. Then, there exists an Abelian anyon $\tilde{v}$ that satisfies $v=\tilde{v}^n$, by using an integer $m$ such that $mn=1 \mod  N_v$ and setting $\tilde{v}=v^m$. Then, $\zeta_n$ is expressed by
\begin{align}
    \begin{split}
        {\sum_{a\in\mathcal{C}}e^{i\pi Q_a} d_a^2\theta_a^n} &= \sum_{a\in\Breve{\mathcal{C}}}\frac{(M_{v,a}+M_{v\times \psi, a})}{2}d_a^2\theta_a^n \\
        &= \sum_{a\in\Breve{\mathcal{C}}}\frac{(M_{\tilde{v}^{n},a}+M_{(\tilde{v}\times \psi)^{n}, a})}{2}d_a^2\theta_a^n \\
        &= \frac{(\theta_{\tilde{v}})^{-n}+(\theta_{\tilde{v}\times \psi})^{-n}}{2}\sum_{a\in\Breve{\mathcal{C}}}d_a^2\theta_a^n \\
        &= (\theta_{\tilde{v}})^{-n}\sum_{a\in\Breve{\mathcal{C}}}d_a^2\theta_a^n \\
        &= (\theta_{v})^{-nm^2}\sum_{a\in\Breve{\mathcal{C}}}d_a^2\theta_a^n
    \end{split}
\end{align}
When a spin TQFT admits a U(1) symmetric gapped boundary, we must have $\theta_v=1$ as shown in Sec.~\ref{sec:U1f}. Also, the presence of gapped boundary in spin TQFT implies that a bosonic theory $\exC\boxtimes D(\Z_2)$ has a Lagrangian algebra anyon, as shown in Sec.~\ref{sec:DZ2}.
Now, $\sum_{a\in\Breve{\mathcal{C}}}d_a^2\theta_a^n$ is higher central charge of a bosonic modular theory $\exC$, and must have a trivial phase for all $n$ with $\gcd(n,N_{\mathrm{FS}}(\Breve{\mathcal{C}}))=1$, when 
$\Breve{\mathcal{C}}\boxtimes D(\Z_2)$ has a Lagrangian algebra anyon~\cite{kaidi2021higher}.
Here, $N_{\mathrm{FS}}(\Breve{\mathcal{C}})$ is Frobenius-Schur exponent of $\Breve{\mathcal{C}}$. So, we have shown that $\zeta_n=1$ for all $n$ with $\gcd(n,N_{\mathrm{FS}}(\Breve{\mathcal{C}}))=1$.

Finally, we show that $\gcd(n,N_{\mathrm{FS}}(\Breve{\mathcal{C}}))=1$ is equivalent to $\gcd(n,N_{\mathrm{FS}}({\mathcal{C}}))=1$. 
This can be checked as follows. Since $v^2\in\mathcal{B}$, $N_v$ is expressed in the form of $N_v=2^{r}\times q$ with $r\ge 1$ and odd number $q$. Consider $x := v^{2^{r}}\in\mathcal{B}$ which has order $q$. Then, there exists an anyon $y\in\mathcal{B}$ such that $y^{2^r}=x$, given by preparing an integer $m$ such that $m\cdot 2^{r}=1$ mod $q$ and setting $y:=x^m\in\mathcal{B}$.
Then we can define $v_0:= v\times y^*\in\exC_{\R}$, which satisfies $v_0^{2^r}=1$. Then, all anyons $x\in \exC_\R$ is expressed as $x=a\times v_0$ with some $a\in\mathcal{C}$, so $\theta_x=\theta_a\theta_{v_0} M_{a,v_0}$. $N_{\mathrm{FS}}(\Breve{\mathcal{C}})$ is then expressed as $N_{\mathrm{FS}}(\Breve{\mathcal{C}})=N_{\mathrm{FS}}(\mathcal{C})\cdot 2^t$ for some integer $t\ge 0$. Since $N_{\mathrm{FS}}(\mathcal{C})$ is even, the condition $\gcd(n,N_{\mathrm{FS}}(\Breve{\mathcal{C}}))=1$ is equivalent to $\gcd(n,N_{\mathrm{FS}}({\mathcal{C}}))=1$. 
\end{proof}

Before closing this section, let us explicitly check with an example that the quantity $\zeta_n$ gives an obstruction beyond the chiral central charge $c_-$ and Hall conductivity $\sigma_H$ given by Eq.~\eqref{eq:c-spinc} and Eq.~\eqref{eq:sigmaspinc}. The simplest example that demonstrates this is a spin TQFT described by a super-modular category $\mathcal{C} = \U_2\times \U_{-4}\times \{1,\psi\}$, where $\psi$ is a transparent fermion.
Let us define $\U^f$ symmetry fractionalization by picking up a vison $v$ in R sector, such that $v$ is a boson. For example, we take the R sector as $\Breve{\mathcal{C}}_{\mathrm{R}}=\U_2\times \U_{-4}\times \{e,m\}$ with $e,m$ bosons in the $\Z_2$ toric code $D(\Z_2)=\{1,\psi,e,m\}$, and then define $v=e$.
Then, one can see that both $c_-$ and $\sigma_H$ are trivial: $e^{2\pi i c_-}=e^{2\pi i \sigma_H}=1$, while the higher version $\zeta_n$ is nontrivial, e.g., $\zeta_3=-1$, reflecting that higher central charge $\xi_3$ of a bosonic modular extension $\Breve{\mathcal{C}} = U(1)_2\times U(1)_{-4}\times D(\Z_2)$ is a nontrivial phase.

\section{Discussions}
\label{sec:discussion}

In this work, we present a general framework for gapped boundary of (2+1)D topological phases with global symmetry, both in the bosonic or fermionic cases. We then derived obstructions to symmetry-preserving gapped boundary, mainly focusing on U(1)-symmetric bosonic and U(1)$^f$-symmetric fermionic phases. 
Let us comment on several generalizations and future questions. Some of them are investigated in future work.

First, for a (2+1)D bosonic topological phase, we only considered a bosonic gapped boundary obtained by condensing bosons. However, it is also possible to obtain a gapped boundary of a bosonic phase by fermion condensation, which gives a fermionic gapped boundary condition~\cite{bhardwajGaiottoKapustin2017, ChenjieWang2017}. For example,~\cite{bhardwajGaiottoKapustin2017} constructed a gapped boundary of a (2+1)D toric code $D(\Z_2)$ by condensing a fermion $\psi$, by introducing fermionic degrees of freedom on the boundary. It would be interesting to make a generalization of the existing algebraic formalism of anyon condensation to the case for fermionic gapped boundary of a bosonic phase. 

Also, it is a natural question to consider symmetry enrichment of such a fermionic gapped boundary. For example, consider a $\Z_2$ global symmetry of $D(\Z_2)$ that permutes anyons as $e\leftrightarrow m$. Then, we immediately find that a bosonic gapped boundary cannot preserve the $\Z_2$ global symmetry, since condensing $e$ or $m$ explicitly violates the symmetry. Meanwhile, condensing $\psi$ is left invariant under this $\Z_2$ symmetry action. So, it is expected that fermion condensation makes possible to obtain richer class of symmetry-preserving gapped boundary for bosonic phases.

In addition, in the study of gapped boundary with $G$ global symmetry for bosonic phases, we only considered the Lagrangian algebra anyon that corresponds to a tube of gapped boundary without $G$ holonomy along its meridian. We can also consider the object $\mathcal{L}_{\mathbf{g}}$ that is given by shrinking the tube of gapped boundary with holonomy labeled by $\mathbf{g}\in G$, which should be an anyon in the twisted sector $\C_{\mathbf{g}}$ of $G$-crossed modular tensor category $\mathcal{C}_G^{\times}$~\cite{barkeshli2019}. 
This should define the $G$-crossed version of the Lagrangian algebra anyon of $\mathcal{C}_G^{\times}$ in the form of 
\begin{align}
    \mathcal{L}_G^{\times}:=\bigoplus_{\mathbf{g}\in G} \mathcal{L}_{\mathbf{g}}.
\end{align}
It would be interesting to study this $G$-crossed Lagrangian algebra anyon in $\mathcal{C}_{G}^{\times}$, and obtain a further constraint on the data of $G$-crossed category required for symmetry-preserving gapped boundary. 

As a related work, Ref.~\cite{Heinrich2018} comments on an expectation about the algebraic criteria for symmetry-preserving gapped boundary of a (2+1)D symmetry-enriched topological phase, based on the Lagrangian algebra of the theory obtained by gauging the global symmetry of the topological phase. Concretely, they conjecture that the existence of the symmetry-preserving gapped boundary of $G$-symmetric topological phase $\mathcal{C}$ is equivalent to the existence of the Lagrangian algebra anyon $(\mathcal{L}_G^{\times})^G$ in a $G$-gauged theory $(\mathcal{C}_G^{\times})^G$, such that for each $\mathbf{g}\in G$ there is at least one condensed anyon $a_\mathbf{g}$ satisfying $\mathrm{Hom}(a_{\mathbf{g}},(\mathcal{L}_G^{\times})^G)\neq 0$ that carries the $[\mathbf{g}]$ magnetic flux, where $[\mathbf{g}]$ is the conjugacy class that contains $\mathbf{g}\in G$.
Though we do not figure out how to show that the above Lagrangian algebra is sufficient for the existence of the symmetry-preserving gapped boundary, we believe that $(\mathcal{L}_G^{\times})^G$ containing the magnetic flux for each conjugacy class $[\mathbf{g}]$ is at least necessary for having a symmetry-preserving gapped boundary.
This is because one can construct the $G$-crossed Lagrangian algebra anyon $\mathcal{L}_G^{\times}$ of the $G$-crossed theory $\mathcal{C}_G^{\times}$ as explained above, which is expected to contain all the nontrivial twisted sector $\mathbf{g}\in G$. Then, we expect that there should be a way to obtain the Lagrangian algebra $(\mathcal{L}_G^{\times})^G$ of the gauged theory, based on the object $\mathcal{L}_G^{\times}$ of the $G$-crossed theory, which should be phrased as $G$-equivariantization of the Lagrangian algebra~\cite{barkeshli2019}. It would be interesting to give an explicit proof to these expectations.

Finally, though we mainly focused on application to U(1) global symmetry, it would be interesting to construct the formula for obstructions to gapped boundary for other symmetries. In particular, for fermionic phases without any global symmetry, one should be able to compute $c_-$ mod 1/2 for a given super-modular category. However, even a formula of $c_-$ mod 1/2 for a super-modular category is not known. So, it would be nice to express any obstruction to gapped boundary of fermionic phases without symmetry in terms of the data of super-modular category.

\section*{Acknowledgement}
We thank Maissam Barkeshli, Kantaro Ohmori and Sahand Seifnashri for useful discussions. We thank Kansei Inamura, Kantaro Ohmori, Yuji Tachikawa and Yunqin Zheng for comments on a draft.
RK is supported by the JQI postdoctoral fellowship at the University of Maryland.

\appendix 

\section{Review of (2+1)D anyon systems}   
\label{app:anyon}
\subsection{Notations of BTC}

In this appendix, we briefly review the notation that we use to describe braided tensor category (BTC). For a more comprehensive review of the notation that we use, see, e.g., Ref.~\cite{barkeshli2019, Kitaevanyons}. The topologically  non-trivial quasiparticles of a (2+1)D topologically ordered state are referred to as anyons. In the category theory terminology, they correspond to isomorphism classes of simple objects of the BTC. 

A BTC $\mathcal{C}$ contains splitting spaces $V_{c}^{ab}$, and their dual fusion spaces, $V_{ab}^c$, where $a,b,c \in \mathcal{C}$ are anyons. These spaces have dimension 
$\text{dim } V_{c}^{ab} = \text{dim } V_{ab}^c = N_{ab}^c$, where the fusion coefficients $N_{ab}^c$ determine the fusion rules. In particular, the fusion rules of the anyons are written as $a \times b = \sum_c N_{ab}^c c$, so that fusion from $a \times b \to c$ is possible if and only if $N_{ab}^c \ge 1$. If $N_{ab}^c > 1$, then each fusion corresponds to a higher dimensional vector space with more possible ``fusion outcomes''. 

The fusion spaces are depicted graphically as: 
\vspace{-30pt}
\begin{equation}
\left( d_{c} / d_{a}d_{b} \right) ^{1/4}
\begin{pspicture}[shift=0.5](-0.1,-0.2)(1.5,1.2)
  \small
  \psset{linewidth=0.9pt,linecolor=black,arrowscale=1.5,arrowinset=0.15}
  \psline{-<}(0.7,0)(0.7,-0.35)
  \psline(0.7,0)(0.7,-0.55)
  \psline(0.7,-0.55) (0.25,-1)
  \psline{-<}(0.7,-0.55)(0.35,-0.9)
  \psline(0.7,-0.55) (1.15,-1)	
  \psline{-<}(0.7,-0.55)(1.05,-0.9)
  \rput[tl]{0}(0.4,0){$c$}
  \rput[br]{0}(1.4,-0.95){$b$}
  \rput[bl]{0}(0,-0.95){$a$}
 \scriptsize
  \rput[bl]{0}(0.85,-0.5){$\mu$}
  \end{pspicture}
=\left\langle a,b;c,\mu \right| \in
V_{ab}^{c} ,
\label{eq:bra}
\end{equation}

\begin{equation}
\left( d_{c} / d_{a}d_{b}\right) ^{1/4}
\begin{pspicture}[shift=-0.65](-0.1,-0.2)(1.5,1.2)
  \small
  \psset{linewidth=0.9pt,linecolor=black,arrowscale=1.5,arrowinset=0.15}
  \psline{->}(0.7,0)(0.7,0.45)
  \psline(0.7,0)(0.7,0.55)
  \psline(0.7,0.55) (0.25,1)
  \psline{->}(0.7,0.55)(0.3,0.95)
  \psline(0.7,0.55) (1.15,1)	
  \psline{->}(0.7,0.55)(1.1,0.95)
  \rput[bl]{0}(0.4,0){$c$}
  \rput[br]{0}(1.4,0.8){$b$}
  \rput[bl]{0}(0,0.8){$a$}
 \scriptsize
  \rput[bl]{0}(0.85,0.35){$\mu$}
  \end{pspicture}
=\left| a,b;c,\mu \right\rangle \in
V_{c}^{ab},
\label{eq:ket}
\end{equation}
where $\mu=1,\ldots ,N_{ab}^{c}$, $d_a$ is the quantum dimension of $a$, 
and the factors $\left(\frac{d_c}{d_a d_b}\right)^{1/4}$ are a normalization convention for the diagrams. 

Diagrammatically, inner products come from connecting the fusion/splitting spaces' lines as:
\begin{equation}
  \begin{pspicture}[shift=-0.95](-0.2,-0.35)(1.2,1.75)
  \small
  \psarc[linewidth=0.9pt,linecolor=black,border=0pt] (0.8,0.7){0.4}{120}{240}
  \psarc[linewidth=0.9pt,linecolor=black,arrows=<-,arrowscale=1.4,
    arrowinset=0.15] (0.8,0.7){0.4}{165}{240}
  \psarc[linewidth=0.9pt,linecolor=black,border=0pt] (0.4,0.7){0.4}{-60}{60}
  \psarc[linewidth=0.9pt,linecolor=black,arrows=->,arrowscale=1.4,
    arrowinset=0.15] (0.4,0.7){0.4}{-60}{15}
  \psset{linewidth=0.9pt,linecolor=black,arrowscale=1.5,arrowinset=0.15}
  \psline(0.6,1.05)(0.6,1.55)
  \psline{->}(0.6,1.05)(0.6,1.45)
  \psline(0.6,-0.15)(0.6,0.35)
  \psline{->}(0.6,-0.15)(0.6,0.25)
  \rput[bl]{0}(0.07,0.55){$a$}
  \rput[bl]{0}(0.94,0.55){$b$}
  \rput[bl]{0}(0.26,1.25){$c$}
  \rput[bl]{0}(0.24,-0.05){$c'$}
 \scriptsize
  \rput[bl]{0}(0.7,1.05){$\mu$}
  \rput[bl]{0}(0.7,0.15){$\mu'$}
  \endpspicture
=\delta _{c c ^{\prime }}\delta _{\mu \mu ^{\prime }} \sqrt{\frac{d_{a}d_{b}}{d_{c}}}
  \pspicture[shift=-0.95](0.15,-0.35)(0.8,1.75)
  \small
  \psset{linewidth=0.9pt,linecolor=black,arrowscale=1.5,arrowinset=0.15}
  \psline(0.6,-0.15)(0.6,1.55)
  \psline{->}(0.6,-0.15)(0.6,0.85)
  \rput[bl]{0}(0.75,1.25){$c$}
  \end{pspicture}
  ,
\end{equation}
This is a way of phrasing topological charge conservation. In addition, we have the usual `resolution of the identity' in a UMTC, phrased diagrammatically:
\begin{equation} \label{resOfIdentity}
\begin{pspicture}[shift=-0.65](-0.1,-0.2)(1.0,1.2)
  \small
  \psset{linewidth=0.9pt,linecolor=black,arrowscale=1.5,arrowinset=0.15}
  \psline{->}(0.25,0)(0.25,0.6)
  \psline(0.25,0)(0.25,1.0)
  \psline{->}(0.7,0)(0.7,0.6)
  \psline(0.7,0)(0.7,1.0)
  \rput[br]{0}(0.15,0.5){$a$}
  \rput[bl]{0}(0.8,0.5){$b$}
 \end{pspicture}
=\sum_{c} \sqrt{\frac{d_c}{d_a d_b}}
\begin{pspicture}[shift=-0.65](-0.4,-0.2)(1.5,1.3)
  \small
 \psset{linewidth=0.9pt,linecolor=black,arrowscale=1.5,arrowinset=0.15}
  \psline{->}(0.7,0.25)(0.7,0.7)
  \psline(0.7,0.25)(0.7,0.8)
  \psline(0.7,0.8) (0.25,1.25)
  \psline{->}(0.7,0.8)(0.3,1.2)
  \psline(0.7,0.8) (1.15,1.25)	
  \psline{->}(0.7,0.8)(1.1,1.2)
  \psline{->}(0.25,-0.3)(0.6,0.15)
  \psline(0.25,-0.3)(0.7,0.25)
  \psline{->}(1.15,-0.3)(0.8,0.15)
  \psline(1.15,-0.3)(0.7,0.25)
  \rput[bl]{0}(0.4,0.5){$c$}
  \rput[br]{0}(1.4,1.05){$b$}
  \rput[bl]{0}(0,1.05){$a$}
  \rput[bl]{0}(0,-0.2){$a$}
  \rput[br]{0}(1.4,-0.2){$b$}
  \end{pspicture},
\end{equation}
implicitly assuming $N_{ab}^c \leq 1$ for all $a,b,c$. 

We denote $\bar{a}$ as the topological charge conjugate of $a$, for which
$N_{a \bar{a}}^1 = 1$, i.e.
\begin{align}
a \times \bar{a} = 1 +\cdots
\end{align}
Here $1$ refers to the identity particle, i.e. the vacuum topological sector, which physically describes all 
local, topologically trivial bosonic excitations. 

The $F$-symbols are defined as the following basis transformation between the splitting
spaces of $4$ anyons:
\begin{equation}
\begin{pspicture}[shift=*](0,-0.45)(1.8,1.8)
  \small
  \psset{linewidth=0.9pt,linecolor=black,arrowscale=1.5,arrowinset=0.15}
  \psline(0.2,1.5)(1,0.5)
  \psline(1,0.5)(1,0)
  \psline(1.8,1.5) (1,0.5)
  \psline(0.6,1) (1,1.5)
  \psline{->}(0.6,1)(0.3,1.375)
  \psline{->}(0.6,1)(0.9,1.375)
  \psline{->}(1,0.5)(1.7,1.375)
  \psline{->}(1,0.5)(0.7,0.875)
  \psline{->}(1,0)(1,0.375)
  \rput[bl]{0}(0.05,1.6){$a$}
  \rput[bl]{0}(0.95,1.6){$b$}
  \rput[bl]{0}(1.75,1.6){${c}$}
  \rput[bl]{0}(0.5,0.5){$e$}
  \rput[bl]{0}(0.9,-0.3){$d$}
 \scriptsize
  \rput[bl]{0}(0.3,0.8){$\alpha$}
  \rput[bl]{0}(0.7,0.25){$\beta$}
\end{pspicture}
= \sum_{f,\mu,\nu} \left[F_d^{abc}\right]_{(e,\alpha,\beta)(f,\mu,\nu)}
 \begin{pspicture}[shift=-1.0](0,-0.45)(1.8,1.8)
  \small
  \psset{linewidth=0.9pt,linecolor=black,arrowscale=1.5,arrowinset=0.15}
  \psline(0.2,1.5)(1,0.5)
  \psline(1,0.5)(1,0)
  \psline(1.8,1.5) (1,0.5)
  \psline(1.4,1) (1,1.5)
  \psline{->}(0.6,1)(0.3,1.375)
  \psline{->}(1.4,1)(1.1,1.375)
  \psline{->}(1,0.5)(1.7,1.375)
  \psline{->}(1,0.5)(1.3,0.875)
  \psline{->}(1,0)(1,0.375)
  \rput[bl]{0}(0.05,1.6){$a$}
  \rput[bl]{0}(0.95,1.6){$b$}
  \rput[bl]{0}(1.75,1.6){${c}$}
  \rput[bl]{0}(1.25,0.45){$f$}
  \rput[bl]{0}(0.9,-0.3){$d$}
 \scriptsize
  \rput[bl]{0}(1.5,0.8){$\mu$}
  \rput[bl]{0}(0.7,0.25){$\nu$}
  \end{pspicture}
.
\end{equation}

To describe topological phases, these are required to be unitary transformations, i.e.

\begin{eqnarray}
\left[ \left( F_{d}^{abc}\right) ^{-1}\right] _{\left( f,\mu
,\nu \right) \left( e,\alpha ,\beta \right) }
&= \left[ \left( F_{d}^{abc}\right) ^{\dagger }\right]
  _{\left( f,\mu ,\nu \right) \left( e,\alpha ,\beta \right) }
  \nonumber \\
&= \left[ F_{d}^{abc}\right] _{\left( e,\alpha ,\beta \right) \left( f,\mu
,\nu \right) }^{\ast }
.
\end{eqnarray}

The $R$-symbols define the braiding properties of the anyons, and are defined via the the following
diagram:
\begin{equation}
\begin{pspicture}[shift=-0.65](-0.1,-0.2)(1.5,1.2)
  \small
  \psset{linewidth=0.9pt,linecolor=black,arrowscale=1.5,arrowinset=0.15}
  \psline{->}(0.7,0)(0.7,0.43)
  \psline(0.7,0)(0.7,0.5)
 \psarc(0.8,0.6732051){0.2}{120}{240}
 \psarc(0.6,0.6732051){0.2}{-60}{35}
  \psline (0.6134,0.896410)(0.267,1.09641)
  \psline{->}(0.6134,0.896410)(0.35359,1.04641)
  \psline(0.7,0.846410) (1.1330,1.096410)	
  \psline{->}(0.7,0.846410)(1.04641,1.04641)
  \rput[bl]{0}(0.4,0){$c$}
  \rput[br]{0}(1.35,0.85){$b$}
  \rput[bl]{0}(0.05,0.85){$a$}
 \scriptsize
  \rput[bl]{0}(0.82,0.35){$\mu$}
  \end{pspicture}
=\sum\limits_{\nu }\left[ R_{c}^{ab}\right] _{\mu \nu}
\begin{pspicture}[shift=-0.65](-0.1,-0.2)(1.5,1.2)
  \small
  \psset{linewidth=0.9pt,linecolor=black,arrowscale=1.5,arrowinset=0.15}
  \psline{->}(0.7,0)(0.7,0.45)
  \psline(0.7,0)(0.7,0.55)
  \psline(0.7,0.55) (0.25,1)
  \psline{->}(0.7,0.55)(0.3,0.95)
  \psline(0.7,0.55) (1.15,1)	
  \psline{->}(0.7,0.55)(1.1,0.95)
  \rput[bl]{0}(0.4,0){$c$}
  \rput[br]{0}(1.4,0.8){$b$}
  \rput[bl]{0}(0,0.8){$a$}
 \scriptsize
  \rput[bl]{0}(0.82,0.37){$\nu$}
  \end{pspicture}
  .
\end{equation}
Under a basis transformation, $\Gamma^{ab}_c : V^{ab}_c \rightarrow V^{ab}_c$, the $F$ and $R$ symbols change:
\begin{align} \label{eq:vertexBasisTransformation}
  F^{abc}_{def} &\rightarrow \check{F}^{abc}_d = \Gamma^{ab}_e \Gamma^{ec}_d F^{abc}_{def} [\Gamma^{bc}_f]^\dagger [\Gamma^{af}_d]^\dagger
  \nonumber \\
  R^{ab}_c & \rightarrow \check{R}^{ab}_c = \Gamma^{ba}_c R^{ab}_c [\Gamma^{ab}_c]^\dagger .
\end{align}
  where we have suppressed splitting space indices and dropped brackets on the $F$-symbol for shorthand.
  These basis transformations are referred to as vertex basis gauge transformations. Physical quantities correspond to gauge-invariant combinations
  of the data. 


The topological twist $\theta_a$ is defined via the diagram:
\begin{equation}
\theta _{a}=\theta _{\bar{a}}
=\sum\limits_{c,\mu } \frac{d_{c}}{d_{a}}\left[ R_{c}^{aa}\right] _{\mu \mu }
= \frac{1}{d_{a}}
\begin{pspicture}[shift=-0.5](-1.3,-0.6)(1.3,0.6)
\small
  \psset{linewidth=0.9pt,linecolor=black,arrowscale=1.5,arrowinset=0.15}
  \psarc[linewidth=0.9pt,linecolor=black] (0.7071,0.0){0.5}{-135}{135}
  \psarc[linewidth=0.9pt,linecolor=black] (-0.7071,0.0){0.5}{45}{315}
  \psline(-0.3536,0.3536)(0.3536,-0.3536)
  \psline[border=2.3pt](-0.3536,-0.3536)(0.3536,0.3536)
  \psline[border=2.3pt]{->}(-0.3536,-0.3536)(0.0,0.0)
  \rput[bl]{0}(-0.2,-0.5){$a$}
  \end{pspicture}
.
\end{equation}
Finally, the modular, or topological, $S$-matrix, is defined as
\begin{equation}
S_{ab} =\mathcal{D}^{-1}\sum
\limits_{c}N_{\bar{a} b}^{c}\frac{\theta _{c}}{\theta _{{a}}\theta _{b}}d_{c}
=\frac{1}{\mathcal{D}}
\begin{pspicture}[shift=-0.4](0.0,0.2)(2.6,1.3)
\small
  \psarc[linewidth=0.9pt,linecolor=black,arrows=<-,arrowscale=1.5,arrowinset=0.15] (1.6,0.7){0.5}{167}{373}
  \psarc[linewidth=0.9pt,linecolor=black,border=3pt,arrows=<-,arrowscale=1.5,arrowinset=0.15] (0.9,0.7){0.5}{167}{373}
  \psarc[linewidth=0.9pt,linecolor=black] (0.9,0.7){0.5}{0}{180}
  \psarc[linewidth=0.9pt,linecolor=black,border=3pt] (1.6,0.7){0.5}{45}{150}
  \psarc[linewidth=0.9pt,linecolor=black] (1.6,0.7){0.5}{0}{50}
  \psarc[linewidth=0.9pt,linecolor=black] (1.6,0.7){0.5}{145}{180}
  \rput[bl]{0}(0.1,0.45){$a$}
  \rput[bl]{0}(0.8,0.45){$b$}
  \end{pspicture}
,
\label{eqn:mtcs}
\end{equation}
where $\mathcal{D} = \sqrt{\sum_a d_a^2}$.

We also denote by $\mathcal{A}$ the Abelian group corresponding to fusion of Abelian anyons, for which each $a \in \mathcal{A}$ satisfies $d_a = 1$ and $a \times b$ has a unique fusion product for any $b \in \mathcal{C}$.

 The double braid, or mutual statistics, of anyons $a$ and $b$ is defined as
 \begin{align}
     M_{ab}=\frac{S^*_{ab}S_{00}}{S_{0a}S_{0b}}
     \label{doubleBraid}
 \end{align}
 and is a phase if either $a$ or $b$ is an Abelian anyon.

\subsection{ Modular and super-modular tensor categories} \label{app:superModularCategories}

        Physically realizable bosonic topological orders are described by modular tensor categories, which have the property that the $S$-matrix is unitary. This means that braiding is non-degenerate, that is, every anyon $a$ can be detected by its non-trivial mutual statistics with some other anyon $b$.
        
        Fermionic topological phases have local fermions, and locality requires that these fermions have trivial mutual statistics with all other excitations. One way to keep track of the anyon fusion and braiding properties in a fermionic topological phase is to use a super-modular tensor category $\mathcal{C}$. A super-modular tensor category is a unitary braided fusion category where there are exactly two transparent anyons: the identity $1$ and a fermion $\psi$. That is, $\psi$ has $\theta_{\psi}=-1$ and trivial mutual braiding with all other particles. As such, the braiding is degenerate.
        
        In a super-modular tensor category, the anyons (simple objects) \it as a set \rm form the structure $\{1,a,b, ...\} \times \{1,\psi\}$.
        The $S$ matrix factorizes as
        \begin{align}
          S = \tilde{S} \otimes \frac{1}{\sqrt{2}} \left(\begin{matrix} 1 & 1 \\ 1 & 1 \end{matrix} \right) ,
        \end{align}
        where $\tilde{S}$ is unitary.

\subsection{Modular extension of super-modular category}

For a given super-modular category $\C$, its modular extension is a modular tensor category $\exC$ that contains $\C$ as a subcategory. In particular, a modular extension $\exC$ is called a minimal modular extension if its quantum dimension satisfies $\mathcal{D}_{\exC}^2 = 2\mathcal{D}_{\C}^2$. It was shown in~\cite{TJF2021minimal} that every super-modular category admits a minimal modular extension. In this paper, we write the minimal modular extension as $\exC=\exC_{\NS}\oplus \exC_\R$, with $\exC_{\NS}=\C$.

We have $M_{a,\psi}=1$ for $a\in\exC_\NS$ since $\exC_\NS=\C$ is super-modular. Meanwhile, using modularity of $\exC$, it follows that $M_{a,\psi}=-1$ for $a\in \exC_\R$, which can be checked by computing  $\sum_{a\in\exC}S^*_{a,\psi} S_{a,1}$ which must be zero since $S$ is unitary,
\begin{align}
\begin{split}
    \sum_{a\in\exC}S^*_{a,\psi} S_{a,1}&=\frac{1}{\mathcal{D}_{\exC}^2}\left( \sum_{a\in\exC_\NS} d_a^2M_{a,\psi} + \sum_{a\in\exC_\R} d_a^2M_{a,\psi} \right) \\
    &=  \frac{1}{2}+\frac{1}{\mathcal{D}_{\exC}^2}\sum_{a\in\exC_\R} d_a^2M_{a,\psi} \\
    &\ge \frac{1}{2} +\frac{1}{\mathcal{D}_{\exC}^2}\sum_{a\in\exC_\R} d_a^2\cdot (-1)=0,
    \end{split}
\end{align}
where we used $M_{a,\psi}=\pm 1$ in the inequality. So, $\sum_{a\in\exC}S^*_{a,\psi} S_{a,1}=0$ is satisfied only when $M_{a,\psi}=-1$ for all $a\in \exC_\R$.

For a given super-modular category $\C$, the choice of a minimal modular extension $\exC$ is not unique. In particular, for the most trivial super-modular category $\C=\{1,\psi\}$, there exists sixteen minimal modular extensions which we denote as $\exC^{\nu}$ for $\nu\in\Z_{16}$~\cite{Kitaevanyons}. Then, for a generic super-modular category $\C$ and its minimal modular extension $\exC$, we can construct sixteen distinct minimal modular extensions of $\C$ by considering $\exC\boxtimes\exC^{\nu}$, and then condensing a pair of fermions $(\psi,\psi)$. This property for sixteen possible choices of minimal modular extensions is called ``sixteen-fold way''~\cite{bruillard2017a}. 

\section{Review on global symmetry of (2+1)D topological phases}

\label{app:symfrac}

In this appendix, we provide a review of symmetry fractionalization of braided tensor categories (BTC), summarizing results from \cite{barkeshli2019,bulmashSymmFrac,aasen21ferm}. We first recall the topological symmetries of BTC, and then describe the symmetry fractionalization of bosonic and fermionic topological phases. We restrict ourselves to unitary global symmetry for simplicity.

\subsection{ Topological symmetry and braided auto-equivalence}

An important property of a BTC $\mathcal{C}$ is the group of ``topological symmetries,'' which are related to ``braided auto-equivalences'' in the mathematical literature. 
The topological symmetries consist of the invertible maps
\begin{align}
\varphi: \mathcal{C} \rightarrow \mathcal{C} .
\end{align}
We consider the different $\varphi$ modulo equivalences known as natural isomorphisms of the form
\begin{align}
    \Upsilon (\ket{a,b;c}) = \frac{\gamma_a\gamma_b}{\gamma_c}\ket{a,b;c}
\end{align}
The equivalence classes $[\varphi]$ then constitute a group, which we denote as Aut$(\mathcal{C})$ ~\cite{barkeshli2019}.
The maps $\varphi$ may permute the anyons:
\begin{align}
\varphi(a) = a' \in \mathcal{C}, 
\end{align}
subject to the constraint that 
\begin{align}
N_{a'b'}^{c'} = N_{ab}^c,\quad S_{a'b'} = S_{ab}, \quad \theta_{a'} &= \theta_a,
\end{align}
The maps $\varphi$ have a corresponding action on the $F$- and $R$- symbols of the theory, as well as on the fusion and splitting spaces, reviewed below.

\subsection{ Global symmetry and symmetry fractionalization: bosonic case}
\label{globsym}

We now consider a bosonic system which has a unitary global symmetry group $G$. In the bosonic case, the global symmetry acts on the anyons and the topological state space through the action of a group homomorphism
\begin{align}
[\rho] : G \rightarrow \text{Aut}(\mathcal{C}) . 
\end{align}
We use the notation $[\rho_{\bf g}] \in \text{Aut}(\mathcal{C})$ for a specific element ${\bf g} \in G$. The square brackets indicate the equivalence class of symmetry maps related by natural isomorphisms, which we define below. $\rho_{\bf g}$ is thus a representative symmetry map of the equivalence class $[\rho_{\bf g}]$. We use the notation
\begin{align}
\,^{\bf g}a \equiv \rho_{\bf g}(a). 
\end{align}
Each $\rho_{\bf g}$ has a unitary action on the fusion/splitting spaces:
\begin{align}
\rho_{\bf g} : V_{ab}^c \rightarrow V_{\,^{\bf g}a \,^{\bf g}b}^{\,^{\bf g}c} .
\end{align}
We choose a basis $\ket{a,b;c,\mu}$ for $V_{ab}^c$ and write the action of $\rho_{\bf g}$ on the basis states as
\begin{align}
\rho_{\bf g} |a,b;c, \mu\rangle = \sum_{\nu} [U_{\bf g}(\,^{\bf g}a ,
\,^{\bf g}b ; \,^{\bf g}c )]_{\mu\nu} \ket{\,^{\bf g} a, \,^{\bf g} b; \,^{\bf g}c,\nu},
  \label{eqn:rhoStates}
 \end{align}
 and the action of $\rho_{\bf g}$ defined on the rest of the fusion/splitting spaces by linearity.
 Here $U_{\bf g}(\,^{\bf g}a , \,^{\bf g}b ; \,^{\bf g}c ) $ is a $N_{ab}^c \times N_{ab}^c$ matrix.

\begin{figure}
    \centering
    \includegraphics[width=0.6\linewidth]{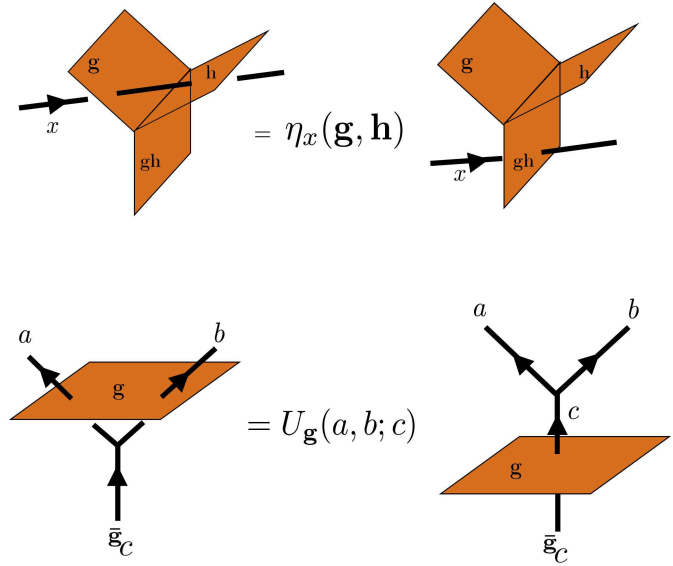}
    \caption{Anyon lines (black) passing through branch sheets (orange) and graphical definitions of the $U$ and $\eta$ symbols. }
        \label{fig:symmFrac}
\end{figure}

In the presence of $G$ symmetry defects, a graphical calculus can be developed for the action of symmetry defects on anyon data. The basic pictures defining the graphical calculus are given in Fig.~\ref{fig:symmFrac}. 
Symmetry fractionalization is specified by a set of phases $\eta_{a}({\bf g,h})$, which satisfy certain consistency relations which we will discuss shortly. The data $\{U,\eta\}$ characterize a symmetry fractionalization class and give us information about how the group symmetries fractionalize onto the different anyons.

There are several consistency conditions that need to be imposed on the $U,\eta,F,R$ symbols in order for diagrammatic evaluations to be consistent under different orders of moves. In the case of $N_{ab}^c \le 1$ they can be written as
\begin{align}
    F^{fcd}_{egl}{F}^{abl}_{efk} = \sum_h {F}^{abc}_{gfh}{F}^{ahd}_{egk}{F}^{bcd}_{khl}
    \label{eqn:Pentagon}
\end{align}
\begin{align}
    {R}^{ca}_e {F}^{acb}_{deg} {R}^{cb}_g = \sum_f {F}^{cab}_{def} {R}^{cf}_d {F}^{abc}_{dfg}
\end{align}
\begin{align}
    ({R}^{ac}_e)^{-1}{F}^{acb}_{deg}({R}^{bc}_g)^{-1} = \sum_f {F}^{cab}_{def}({R}^{fc}_d)^{-1}{F}^{abc}_{dfg}
\end{align}
\begin{align}
    {U}_{\mathbf{g}}(^\mathbf{g}{a},^\mathbf{g}{b} ;^\mathbf{g}{e}){U}_{\mathbf{g}}(^\mathbf{g}{e},^\mathbf{g}{c} ;^\mathbf{g}{d}){F}^{^\mathbf{g}{a}^\mathbf{g}{b}^\mathbf{g}{c}}_{^\mathbf{g}{d}^\mathbf{g}{e}^\mathbf{g}{f}} {U}_{\mathbf{g}}^{-1}(^\mathbf{g}{b},^\mathbf{g}{c} ;^\mathbf{g}{f}){U}_{\mathbf{g}}^{-1}(^\mathbf{g}{a},^\mathbf{g}{f} ;^\mathbf{g}{d})= F^{abc}_{def}
\end{align}
\begin{align}
    {U}_{\mathbf{g}}(^\mathbf{g}{a},^\mathbf{g}{b} ;^\mathbf{g}{c})R^{^\mathbf{g}{a}^\mathbf{g}{b}}_{^\mathbf{g}{c}}{U}_{\mathbf{g}}(^\mathbf{g}{a},^\mathbf{g}{b} ;^\mathbf{g}{c})^{-1}=R^{ab}_c
\end{align}
\begin{align}
    {U}_{\mathbf{g}}({a},{b} ;{c}){U}_{\mathbf{h}}({a},{b} ;{c}) = {U}_{\mathbf{gh}}({a},{b} ;{c})\frac{\eta_c({\bf g}, {\bf h})}{\eta_a({\bf g}, {\bf h})\eta_b({\bf g}, {\bf h})}
\end{align}
\begin{align}
    {\eta}_{a}({\bf h},{\bf k}) {\eta}_a({\bf gh}, {\bf k})= {\eta}_a({\bf g},{\bf h}) {\eta}_a({\bf g}, {\bf hk})
\end{align}
The top three are just the standard pentagon and hexagon equations from BTCs without symmetry. The next two ensure the symmetry action is compatible with the $F$- and $R$-symbols. The next ensures that the symmetry action and symmetry fractionalization are consistent with each other, and the last one is a generalized associativity condition for the $\eta$ symbols.

These data are subject to an additional class of gauge transformations, which arise by changing $\rho$ by a natural isomorphism: \cite{barkeshli2019}
\begin{align}
  U_{\bf g}(a,b;c) &\rightarrow \frac{\gamma_{a}({\bf g}) \gamma_b({\bf g})}{ \gamma_c({\bf g}) } U_{\bf g}(a,b;c)
\nonumber \\
  \eta_a({\bf g}, {\bf h}) & \rightarrow \frac{\gamma_a({\bf g h}) }{\gamma_{\,^{\bf g} a}({\bf h}) \gamma_a({\bf g}) } \eta_a({\bf g}, {\bf h})
  \label{eq:naturaliso}
\end{align}
In this paper we will always fix the gauge
\begin{align}
  \eta_1({\bf g},{\bf h})=\eta_a({\bf 1},{\bf g}) = \eta_a({\bf g},{\bf 1})&=1
                                                                             \nonumber \\
  U_{\bf g}(1,b;c)=U_{\bf g}(a,1;c)&=1.
  \label{eq:fixUeta}
  \end{align}
  
One can show that symmetry fractionalization forms a torsor over $H^2_{\rho}(BG, \mathcal{B})$ in the bosonic case, where $\mathcal{B}$ is a group of Abelian anyons in $\C$. That is, different possible patterns of symmetry fractionalization can be related to each other by elements of ${H}^2_{\rho}(BG, \mathcal{B})$. In particular, given an element $[\mathfrak{t}] \in {H}^2_{\rho}(BG, \mathcal{B})$, we can change the symmetry fractionalization class as
\begin{align}
\eta_a({\bf g}, {\bf h}) \rightarrow \eta_a({\bf g}, {\bf h}) M_{a \mathfrak{t}({\bf g},{\bf h})},
\end{align}
where $\mathfrak{t}({\bf g},{\bf h}) \in \mathcal{B}$ is a representative 2-cocycle for the cohomology class $[\mathfrak{t}]$ and $M_{ab}$ is the mutual braiding in Eq.~\eqref{doubleBraid}.

\subsection{ Fermionic symmetries}
\label{fermSymFracSec}

        We sketch a description of fermionic symmetry in fermionic topological phases where $\C$ is super-modular, summarizing results from~\cite{bulmashSymmFrac, aasen21ferm}. The anyons in a super-modular category $\C$ corresponds to the states in the NS sector of the Hilbert space in spin TQFT, so symmetry fractionalization for $\C$ is understood as specifying the symmetry action on the NS sector~\cite{delmastro2021}. To describe the symmetry action on the whole Hilbert space including the R sector, we need to define symmetry fractionalization on the modular extension $\exC=\exC_\NS\oplus\exC_\R$ of $\C=\exC_\NS$.
        
        As we review in Appendix~\ref{app:spinTQFT}, a spin TQFT is constructed by performing a process of ``fermion condensation'' for a given bosonic TQFT.
        This process corresponds to gauging 1-form symmetry generated by a Wilson line of a transparent fermion $\psi$, and the resulting theory has a dual $\Z_2^f$ fermion parity symmetry. 
        
        Suppose we want a fermionic symmetry with a fermionic group $G_f$ after fermion condensation, where $G_f$ is defined by a symmetry extension $\Z_2^f\to G_f\to G_b$ with the extension characterized by $\omega_2\in Z^2(BG_b,\Z_2)$. Then, the initial bosonic theory (sometimes called a ``bosonic shadow theory'') must have $G_b$ symmetry, where the $G_b$ symmetry has a certain mixed 't Hooft anomaly between the $\Z_2$ 1-form symmetry generated by $\psi$ Wilson line~\cite{Tachikawa:2017gyf, tata2021anomalies}.
        So, we start with constructing a symmetry action of $G_b$ for the bosonic shadow theory described by a modular category $\exC=\exC_\NS\oplus\exC_\R$, taking into account certain additional constraints which arise from the mixed 't Hooft anomaly. 
        
        The $G_b$ action of $\exC$ is realized as a topological symmetry of BTC which preserves the $\Z_2$ grading $\exC=\exC_\NS\oplus\exC_\R$. So, it induces a topological symmetry of the NS sector $\exC_\NS=\C$, which was formulated by~\cite{bulmashSymmFrac, aasen21ferm}. 
        One can immediately see that $G_b$ cannot permute $\psi$; this follows from the super-modularity of $\mathcal{C}$, where the topological symmetry has to preserve the transparent fermion $\psi$.
        
        Then, we need to account for constraints on symmetry fractionalization data which comes from mixed 't Hooft anomaly between $G_b$ symmetry and a $\Z_2$ 1-form symmetry generated by $\psi$ Wilson line. The 't Hooft anomaly is given by a (3+1)D response action $(-1)^{\int A^*\omega\cup B}$, where $A: M\to BG_b$ is a $G_b$ background gauge field and $B\in Z^2(M,\Z_2)$ is a background gauge field of $\Z_2$ 1-form symmetry. This mixed 't Hooft anomaly implies that a fermion $\psi$ carries symmetry fractionalization characterized by $\omega\in Z^2(BG_b,\Z_2)$~\cite{benini2019, tata2021anomalies, thorngren2015framed}.
        
        To realize the desired mixed 't Hooft anomaly, we impose constraints
        \begin{equation}
            U_{\bf g}(\psi, \psi;1)=1
            \label{eqn:Upsi1}
        \end{equation}
        \begin{eqnarray}
        \eta_{\psi}({\bf g,h}) = \omega_2({\bf g,h})
        \label{eqn:etaPsiOmega2Constraint}
        \end{eqnarray}
        These equations together with~\eqref{eq:fixUeta} guarantees that the phase shifted by moving the $\psi$ Wilson line across $G_b$ symmetry defects is entirely characterized by~\eqref{eqn:etaPsiOmega2Constraint}, which correctly realizes the effect of mixed 't Hooft anomaly under gauge transformation of 1-form symmetry.
        
        Physically,~\eqref{eqn:Upsi1} arises from the idea that the symmetry transformation rules of the local fermion operators in a fermionic theory are set entirely by the local Hilbert space, that is, the action of the symmetry on states containing only fermions must be determined entirely by the local action of the symmetry. Thus the action on the topological state space given by $U_{\bf g}(\psi,\psi;1)$ must be trivial. 
        In order to preserve above two constraints, we have a constraint on gauge transformations by a natural isomorphism
        \begin{equation}
            \gamma_{\psi}({\bf g})=1.
            \label{eqn:gammaPsiConstraint}
        \end{equation}

\section{Review of spin TQFT}
\label{app:spinTQFT}
In this appendix, we review the construction of spin TQFT based on a method of fermion condensation, and describe its Hilbert space. The content of this review is mainly based on~\cite{delmastro2021}. To obtain a spin TQFT, we start with a bosonic TQFT described by a minimal modular extension $\exC$ of a super-modular category $\C$, which is called a ``bosonic shadow theory''. We then gauge $\Z_2$ 1-form symmetry generated by a Wilson line of a transparent fermion $\psi$ in $\C$. The resulting theory has a dual $\Z_2^f$ symmetry which is fermion parity symmetry, and gives a fermionic spin TQFT. The process of gauging the 1-form $\Z_2$ symmetry for $\psi$ is called fermion condensation~\cite{Gaiotto:2015zta, aasen2019, bhardwajGaiottoKapustin2017}.


The Hilbert space of the spin TQFT after fermion condensation is described by using the anyons of the initial bosonic shadow theory $\exC$. Let us consider the Hilbert space on a torus equipped with spin structure $T^2_{\mu,\lambda}$, where $\mu$ is spin structure along the meridian and $\lambda$ is that on the longitude. For simplicity, we assume that all anyons $x\in\exC_\R$ satisfy $x\times \psi\neq x$. For the case in the presence of anyons $\sigma\times \psi=\sigma$, see~\cite{delmastro2021}. When all anyons $x\in\exC_\R$ satisfy $x\times \psi\neq x$, the Hilbert space of the spin TQFT is spanned by states described as follows:
\begin{align}
\begin{split}
    \mathcal{H}(T^2_{\NS,\NS}):& \qquad \ket{a} + \ket{a\times\psi}, \quad a\in\exC_{\NS}/\{1,\psi\} \\
    \mathcal{H}(T^2_{\NS,\R}):& \qquad \ket{a} - \ket{a\times\psi}, \quad a\in\exC_{\NS}/\{1,\psi\} \\
    \mathcal{H}(T^2_{\R,\NS}):& \qquad \ket{x} + \ket{x\times\psi}, \quad x\in\exC_{\R}/\{1,\psi\} \\
    \mathcal{H}(T^2_{\R,\R}):& \qquad \ket{x} - \ket{x\times\psi}, \quad x\in\exC_{\R}/\{1,\psi\} \\
\label{eq:spinTQFTH}
\end{split}
\end{align}
where $\ket{a}$ denotes a state given by preparing a solid torus $D^2\times S^1$ with insertion of a Wilson line for the anyon $a$ along $S^1$. The above expression of the state is understood in a following way. First, since we are gauging the 1-form symmetry generated by a Wilson line for $\psi$, each state realized in the gauged theory is averaged over all possible insertions of the $\psi$ Wilson line in the spacetime 3-manifold, so that the state is invariant under insertion of $\psi$ Wilson line along both the meridian and longitude. That explains why the state takes the form of $\ket{a}\pm\ket{a\times\psi}$ for each spin structure.

Second, we note that the action of a $\psi$ Wilson line with trivial framing along a curve $C$ on the state on $T^2$ characterizes spin structure along the curve $C$. For example, the action of the $\psi$ Wilson line along the meridian gives a phase $+1$ (resp.~$-1$), if spin structure along the meridian is $\mu=\NS$ (resp.~$\mu=\R$). Due to $M_{x,\psi}=-1$ when $x\in\exC_\R$, we can see that each state in~\eqref{eq:spinTQFTH} corresponds to the desired spin structure on $T^2$.

Let us explain more precisely how an insertion of the $\psi$ Wilson line amounts to measuring spin structure, by describing fermion condensation in detail. 
In general, fermion condensation in (2+1)D starts with a bosonic shadow theory $Z_b$ with a specific $\Z_2$ 1-form symmetry generated by a fermionic Wilson line $\psi$. Its background gauge field is written as $B\in Z^2(M,\Z_2)$.
We want to gauge the 1-form symmetry by making $B$ dynamical. 
However, reflecting the fermionic statistics of $\psi$, it has a 't Hooft anomaly  whose response action is characterized by
\begin{align}
    (-1)^{\int B\cup B}
    \label{eq:shadowanomaly}
\end{align}
This is  understood as a ``framing anomaly'', which is a phase ambiguity of a correlation function with $\psi$ Wilson line inserted, under the change of the framing along the Wilson line~\cite{tata2021anomalies}. 

So the bosonic shadow theory has a nontrivial 't Hooft anomaly, which prevents us from gauging the 1-form symmetry generated by $\psi$. However, when a spacetime 3-manifold is equipped with a spin structure, the anomaly~\eqref{eq:shadowanomaly} can be trivialized. 
This can be seen by using the Wu relation $B\cup B = w_2\cup B$ mod 2 in cohomology with the second Stiefel-Whitney class $[w_2]\in H^2(M,\Z_2)$~\cite{ManifoldAtlasWu}. Then, the anomaly $w_2\cup B$ can be trivialized by introducing spin structure $\xi\in C^1(M,\Z_2)$ on a 3-manifold, where $\delta\xi=w_2$. This means that one can cancel the anomaly of the bosonic shadow theory~\eqref{eq:shadowanomaly} by introducing a counter-term that depends on spin structure. So, we write this counter-term as $z(\xi, B)$, which becomes a phase and depends on spin structure $\xi$. Then, we can gauge the 1-form symmetry of $Z_b$ coupled with the spin theory $z(\xi, B)$,
\begin{align}
    Z_f(\xi) \propto \sum_{B\in Z^2(M,\Z_2)} Z_b(B)z(\xi, B),
    \label{eq:pathintegralcond}
\end{align}
then we obtain a spin theory $Z_f(\xi)$. This process is called fermion condensation. Then, one can see that the fermion condensation is a sum over the configuration of $\psi$ Wilson lines, weighted by a phase $z(\xi, B)$. 

Basically, the phase $z(\xi, B)$ measures a spin structure along the curve where $\psi$ Wilson line is inserted. Concretely, when we consider $z(\xi, B)$ on a 3-manifold $T^2_{\mu,\lambda}\times\mathbb{R}$, then an insertion of a $\psi$ line along the meridian gives a phase $+1$ (resp.~$-1$) when $\mu=\NS$ (resp.~$\mu=\R$), same for the longitude.~\footnote{Here, we assume that the tangent bundle of $T^2_{\mu,\lambda}\times\mathbb{R}$ is framed in the obvious way in $x,y,z$ directions, and the $\psi$ line in $x$ or $y$ direction carries a trivial framing compared with the background framing on $T^2_{\mu,\lambda}\times\mathbb{R}$. Since the theory $z(\xi,B)$ also carries a framing anomaly, we need to keep track of framing of the $\psi$ Wilson line when we talk about its eigenvalue. See~\cite{Gaiotto:2015zta, tata2020, tata2021anomalies} for detailed explanation of the theory $z(\xi, B)$. } Hence,~\eqref{eq:pathintegralcond} is understood as sum over $\psi$ Wilson lines weighted by spin structure along the curve. That explains why the insertion of a $\psi$ Wilson line measures spin structure.

\section{$S$-matrix action on boundary states of spin TQFT}
\label{app:smatrix}
In this appendix, we describe a detailed discussion of $S$-matrix action on boundary states of spin TQFT. 

Since the boundary state $\ket{\mathcal{L}}$ depends on the spin structure of the torus, the evaluation of the Hopf link is done by cases. We write the spin structure for $\ket{\mathcal{L}}$ as $(\mu,\lambda)$, which denotes the meridian and longitude of $T^2$ respectively. Similarly, the spin structure around the anyon line $x$ linked with $\mathcal{L}$ is written as $(\mu',\lambda')$, which satisfies $(\mu',\lambda')=(\lambda,\mu)$.

\begin{itemize}
    \item When $(\mu,\lambda)=(\mathrm{NS},\mathrm{NS})$, the boundary state becomes $\ket{\mathcal{L}_{\mathrm{NS},\mathrm{NS}}}$. We have $(\mu',\lambda')=(\mathrm{NS},\mathrm{NS})$, and the state on $T^2_{\mu',\lambda'}$ is identified as the line operator $x$ in the form of $x=a+a\times\psi$ with $a\in\exC_{\mathrm{NS}}$. The Hopf link amplitude is given by the $S$ matrix between $\ket{\mathcal{L}_{\mathrm{NS},\mathrm{NS}}}$ and $\ket{x}=\ket{a}+\ket{a\times\psi}$ as
    \begin{align}
        \sum_{b\in\mathcal{A}_{\mathrm{NS}}^1}(S_{ab}+S_{a,b\times\psi}+S_{a\times\psi,b}+S_{a\times\psi,b\times\psi})Z_{0b}^{\mathrm{NS}}=\sum_{b\in\mathcal{A}_{\mathrm{NS}}^1}4S_{ab}Z_{0b}^{\mathrm{NS}}
    \end{align}
    Meanwhile, the Hopf link is regarded as a partition function on $D^2\times S^1_\mathrm{NS}$ with a gapped boundary and insertion of $x$ along $S^1$. This further reduces to a partition function on $S^2\times S^1_\mathrm{NS}$ by shrinking boundary of $D^2$ into a point, with insertion of $x\times \overline{\mathcal{L}}_{\mathrm{NS},\mathrm{NS}}$ (see Fig.~\ref{fig:spheredisk}). This is evaluated as
    \begin{align}
        \bra{\mathcal{L}_{\mathrm{NS},\mathrm{NS}}}(\ket{a}+\ket{a\times\psi})=2Z_{0a}^{\mathrm{NS}}
    \end{align}

By comparing the above two expressions, we obtain
\begin{align}
    \sum_{b\in\mathcal{A}_{\mathrm{NS}}^1}2S_{ab}Z_{0b}^{\mathrm{NS}}=Z_{0a}^{\mathrm{NS}} \quad \text{for $a\in\exC_{\mathrm{NS}}$}.
\end{align}
Since $S_{ab}=S_{a,b\times\psi}$ for $a\in\exC_{\mathrm{NS}}$, we also have
\begin{align}
    \sum_{b\in\mathcal{A}_{\mathrm{NS}}^\psi}2S_{ab}Z_{0b}^{\mathrm{NS}}=Z_{0a}^{\mathrm{NS}} \quad \text{for $a\in\exC_{\mathrm{NS}}$}.
\end{align}

\begin{figure}[htb]
\centering
\includegraphics{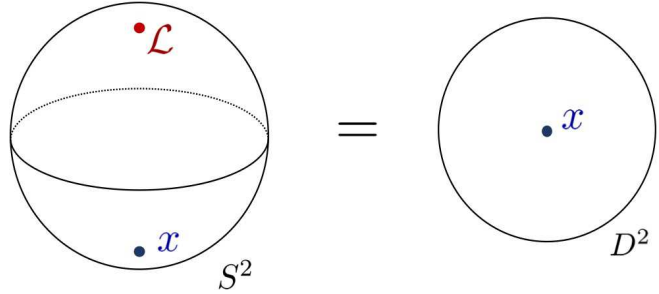}
\caption{A space $D^2$ with a gapped boundary is topologically equivalent to $S^2$ by shrinking a boundary circle into a point.}
\label{fig:spheredisk}
\end{figure}       

\item When $(\mu,\lambda)=(\mathrm{NS},\mathrm{R})$, the boundary state becomes $\ket{\mathcal{L}_{\mathrm{NS},\mathrm{R}}}$. We have $(\mu',\lambda')=(\mathrm{R},\mathrm{NS})$, and the state on $T^2_{\mu',\lambda'}$ is given in the form of $\ket{x}=\ket{a}+\ket{a\times\psi}$ with $a\in\exC_{\mathrm{R}}$. The Hopf link amplitude between them is given by the $S$ matrix as
    \begin{align}
        \sum_{b\in\mathcal{A}_{\mathrm{NS}}^1}(S_{ab}-S_{a,b\times\psi}+S_{a\times\psi,b}-S_{a\times\psi,b\times\psi})Z_{0b}^{\mathrm{NS}}=\sum_{b\in\mathcal{A}_{\mathrm{NS}}^1}4S_{ab}Z_{0b}^{\mathrm{NS}}
    \end{align}
    Meanwhile, the Hopf link is regarded as a partition function on $D^2\times S^1_\mathrm{NS}$ with a gapped boundary and insertion of $x$ along $S^1$. This further reduces to a partition function on $S^2\times S^1_\mathrm{NS}$ by shrinking boundary of $D^2$ into a point, with insertion of $x\times \overline{\mathcal{L}}_{\mathrm{R},\mathrm{NS}}$. This is evaluated as
    \begin{align}
        \bra{\mathcal{L}_{\mathrm{R},\mathrm{NS}}}(\ket{a}+\ket{a\times\psi})=2Z_{0a}^{\mathrm{R}}
    \end{align}
    By comparing the above two expressions, we obtain
    \begin{align}
        \sum_{b\in\mathcal{A}_{\mathrm{NS}}^1}2S_{ab}Z_{0b}^{\mathrm{NS}}=Z_{0a}^{\mathrm{R}} \quad \text{for $a\in\exC_{\mathrm{R}}$}.
    \end{align}
    Since $S_{ab}=-S_{a,b\times\psi}$ for $a\in\exC_{\mathrm{R}}$, we also have
    \begin{align}
        \sum_{b\in\mathcal{A}_{\mathrm{NS}}^\psi}2S_{ab}Z_{0b}^{\mathrm{NS}}=-Z_{0a}^{\mathrm{R}} \quad \text{for $a\in\exC_{\mathrm{R}}$}.
    \end{align}
    
    \item When $(\mu,\lambda)=(\mathrm{R},\mathrm{NS})$, the boundary state becomes $\ket{\mathcal{L}_{\mathrm{R},\mathrm{NS}}}$. We have $(\mu',\lambda')=(\mathrm{NS},\mathrm{R})$, and the state on $T^2_{\mu',\lambda'}$ is given in the form of $\ket{x}=\ket{a}-\ket{a\times\psi}$ with $a\in\exC_{\mathrm{NS}}$. The Hopf link amplitude between them is given by the $S$ matrix as
    \begin{align}
        \sum_{b\in\mathcal{A}_{\mathrm{R}}^e}(S_{ab}+S_{a,b\times\psi}-S_{a\times\psi,b}-S_{a\times\psi,b\times\psi})Z_{0b}^{\mathrm{R}}=\sum_{b\in\mathcal{A}_{\mathrm{R}}^e}4S_{ab}Z_{0b}^{\mathrm{R}}
    \end{align}
    Meanwhile, the Hopf link is regarded as a partition function on $D^2\times S^1_\mathrm{R}$ with a gapped boundary and insertion of $x$ along $S^1$. This further reduces to a partition function on $S^2\times S^1_\mathrm{R}$ by shrinking boundary of $D^2$ into a point, with insertion of $x\times \overline{\mathcal{L}}_{\mathrm{NS},\mathrm{R}}$. This is evaluated as
    \begin{align}
    \begin{split}
        \bra{\mathcal{L}_{\mathrm{NS},\mathrm{R}}}(\ket{a}-\ket{a\times\psi})&=2Z_{0a}^{\mathrm{NS}}\quad \text{if $a\in \mathcal{A}_{\mathrm{NS}}^1$} \\
        \bra{\mathcal{L}_{\mathrm{NS},\mathrm{R}}}(\ket{a}-\ket{a\times\psi})&=-2Z_{0a}^{\mathrm{NS}}\quad \text{if $a\in \mathcal{A}_{\mathrm{NS}}^\psi$} \\
        \bra{\mathcal{L}_{\mathrm{NS},\mathrm{R}}}(\ket{a}-\ket{a\times\psi})&=0\quad \text{if $a\in \exC_{\mathrm{NS}}$ is not an element of $\mathcal{A}_{\mathrm{NS}}$}
        \end{split}
    \end{align}
    By comparing the above two expressions, we obtain the nonzero equations as
    \begin{align}
    \begin{split}
        \sum_{b\in\mathcal{A}_{\mathrm{R}}^e}2S_{ab}Z_{0b}^{\mathrm{R}}&=Z_{0a}^{\mathrm{NS}} \quad \text{if $a\in \mathcal{A}_{\mathrm{NS}}^1$} \\
        \sum_{b\in\mathcal{A}_{\mathrm{R}}^e}2S_{ab}Z_{0b}^{\mathrm{R}}&=-Z_{0a}^{\mathrm{NS}} \quad \text{if $a\in \mathcal{A}_{\mathrm{NS}}^\psi$} \\
        \end{split}
    \end{align}
    We also have
     \begin{align}
    \begin{split}
        \sum_{b\in\mathcal{A}_{\mathrm{R}}^m}2S_{ab}Z_{0b}^{\mathrm{R}}&=Z_{0a}^{\mathrm{NS}} \quad \text{if $a\in \mathcal{A}_{\mathrm{NS}}^1$} \\
        \sum_{b\in\mathcal{A}_{\mathrm{R}}^m}2S_{ab}Z_{0b}^{\mathrm{R}}&=-Z_{0a}^{\mathrm{NS}} \quad \text{if $a\in \mathcal{A}_{\mathrm{NS}}^\psi$} \\
        \end{split}
    \end{align}
    
        \item When $(\mu,\lambda)=(\mathrm{R},\mathrm{R})$, the boundary state becomes $\ket{\mathcal{L}_{\mathrm{R},\mathrm{R}}}$. We have $(\mu',\lambda')=(\mathrm{R},\mathrm{R})$, and the state on $T^2_{\mu',\lambda'}$ is given in the form of $x=a-a\times\psi$ with $a\in\exC_{\mathrm{R}}$. The Hopf link amplitude between them is given by the $S$ matrix as
    \begin{align}
        \sum_{b\in\mathcal{A}_{\mathrm{R}}^e}(S_{ab}-S_{a,b\times\psi}-S_{a\times\psi,b}+S_{a\times\psi,b\times\psi})Z_{0b}^{\mathrm{R}}=\sum_{b\in\mathcal{A}_{\mathrm{R}}^e}4S_{ab}Z_{0b}^{\mathrm{R}}
    \end{align}
    Meanwhile, the Hopf link is regarded as a partition function on $D^2\times S^1_\mathrm{R}$ with a gapped boundary and insertion of $x$ along $S^1$. This further reduces to a partition function on $S^2\times S^1_\mathrm{R}$ by shrinking boundary of $D^2$ into a point, with insertion of $x\times \overline{\mathcal{L}}_{\mathrm{R},\mathrm{R}}$. This is evaluated as
    \begin{align}
    \begin{split}
        \bra{\mathcal{L}_{\mathrm{R},\mathrm{R}}}(\ket{a}-\ket{a\times\psi})&=2Z_{0a}^{\mathrm{R}} \quad \text{if $a\in \mathcal{A}_{\mathrm{R}}^e$}  \\
        \bra{\mathcal{L}_{\mathrm{R},\mathrm{R}}}(\ket{a}-\ket{a\times\psi})&=-2Z_{0a}^{\mathrm{R}} \quad \text{if $a\in \mathcal{A}_{\mathrm{R}}^m$} \\
        \bra{\mathcal{L}_{\mathrm{R},\mathrm{R}}}(\ket{a}-\ket{a\times\psi})&=0 \quad \text{if $a\in \exC_{\mathrm{R}}$ is not an element of $\mathcal{A}_{\mathrm{R}}$} 
        \end{split}
    \end{align}
    By comparing the above two expressions, we obtain the nonzero equations as
    \begin{align}
    \begin{split}
        \sum_{b\in\mathcal{A}_{\mathrm{R}}^e}2S_{ab}Z_{0b}^{\mathrm{R}}&=Z_{0a}^{\mathrm{R}} \quad \text{if $a\in \mathcal{A}_{\mathrm{R}}^e$} \\
        \sum_{b\in\mathcal{A}_{\mathrm{R}}^e}2S_{ab}Z_{0b}^{\mathrm{R}}&=-Z_{0a}^{\mathrm{R}} \quad \text{if $a\in \mathcal{A}_{\mathrm{R}}^m$} \\
        \end{split}
    \end{align}        
    We also have
      \begin{align}
    \begin{split}   
        \sum_{b\in\mathcal{A}_{\mathrm{R}}^m}2S_{ab}Z_{0b}^{\mathrm{R}}&=-Z_{0a}^{\mathrm{R}} \quad \text{if $a\in \mathcal{A}_{\mathrm{R}}^e$} \\
        \sum_{b\in\mathcal{A}_{\mathrm{R}}^m}2S_{ab}Z_{0b}^{\mathrm{R}}&=Z_{0a}^{\mathrm{R}} \quad \text{if $a\in \mathcal{A}_{\mathrm{R}}^m$}
        \end{split}
    \end{align}
\end{itemize}

\bibliographystyle{utphys}
\bibliography{bibliography}{}

\providecommand{\href}[2]{#2}\begingroup\raggedright\begin{thebibliography}{10}

\bibitem{tsui1982}
D.~C. Tsui, H.~L. Stormer, and A.~C. Gossard, ``Two-dimensional
  magnetotransport in the extreme quantum limit,'' {\em Phys. Rev. Lett.}
  {\bfseries 48} (1982) 1559.

\bibitem{laughlin1983}
R.~B. Laughlin, ``Anomalous quantum hall effect: An incompressible quantum
  fluid with fractionally charged excitations,'' {\em Phys. Rev. Lett.}
  {\bfseries 50} (1983) 1395.

\bibitem{wilczek1982}
F.~Wilczek, ``Magnetic flux, angular momentum, and statistics,''
  \href{http://dx.doi.org/10.1103/PhysRevLett.48.1144}{{\em Phys. Rev. Lett.}
  {\bfseries 48} (Apr, 1982) 1144--1146}.
  \url{https://link.aps.org/doi/10.1103/PhysRevLett.48.1144}.

\bibitem{moore1991}
G.~Moore and N.~Read, ``Nonabelions in the fractional quantum hall effect,''
  {\em Nuclear Physics B} {\bfseries 360} no.~2-3, (1991) 362 -- 396.

\bibitem{Banerjee2018}
M.~Banerjee, M.~Heiblum, V.~Umansky, D.~E. Feldman, Y.~Oreg, and A.~Stern,
  ``Observation of half-integer thermal hall conductance,''
  \href{http://dx.doi.org/10.1038/s41586-018-0184-1}{{\em Nature} {\bfseries
  559} no.~7713, (Jun, 2018) 205--210}.

\bibitem{Kitaev:2011dxc}
A.~Kitaev and L.~Kong, ``{Models for Gapped Boundaries and Domain Walls},''
  \href{http://dx.doi.org/10.1007/s00220-012-1500-5}{{\em Commun. Math. Phys.}
  {\bfseries 313} no.~2, (2012) 351--373},
  \href{http://arxiv.org/abs/1104.5047}{{\ttfamily arXiv:1104.5047
  [cond-mat.str-el]}}.

\bibitem{Kapustin:2010hk}
A.~Kapustin and N.~Saulina, ``{Topological boundary conditions in abelian
  Chern-Simons theory},''
  \href{http://dx.doi.org/10.1016/j.nuclphysb.2010.12.017}{{\em Nucl. Phys.}
  {\bfseries B845} (2011) 393--435},
\href{http://arxiv.org/abs/1008.0654}{{\ttfamily arXiv:1008.0654 [hep-th]}}.

\bibitem{Kapustin:2013nva}
A.~Kapustin, ``{Ground-state degeneracy for abelian anyons in the presence of
  gapped boundaries},''
  \href{http://dx.doi.org/10.1103/PhysRevB.89.125307}{{\em Phys. Rev. B}
  {\bfseries 89} no.~12, (2014) 125307},
  \href{http://arxiv.org/abs/1306.4254}{{\ttfamily arXiv:1306.4254
  [cond-mat.str-el]}}.

\bibitem{Barkeshli:2013jaa}
M.~Barkeshli, C.-M. Jian, and X.-L. Qi, ``{Classification of Topological
  Defects in Abelian Topological States},''
  \href{http://dx.doi.org/10.1103/PhysRevB.88.241103}{{\em Phys. Rev. B}
  {\bfseries 88} (2013) 241103},
  \href{http://arxiv.org/abs/1304.7579}{{\ttfamily arXiv:1304.7579
  [cond-mat.str-el]}}.

\bibitem{Hung:2014tba}
L.-Y. Hung and Y.~Wan, ``{Ground State Degeneracy of Topological Phases on Open
  Surfaces},'' \href{http://dx.doi.org/10.1103/PhysRevLett.114.076401}{{\em
  Phys. Rev. Lett.} {\bfseries 114} no.~7, (2015) 076401},
  \href{http://arxiv.org/abs/1408.0014}{{\ttfamily arXiv:1408.0014
  [cond-mat.str-el]}}.

\bibitem{Wang:2012am}
J.~Wang and X.-G. Wen, ``{Boundary Degeneracy of Topological Order},''
  \href{http://dx.doi.org/10.1103/PhysRevB.91.125124}{{\em Phys. Rev. B}
  {\bfseries 91} no.~12, (2015) 125124},
  \href{http://arxiv.org/abs/1212.4863}{{\ttfamily arXiv:1212.4863
  [cond-mat.str-el]}}.

\bibitem{Lan:2014uaa}
T.~Lan, J.~C. Wang, and X.-G. Wen, ``{Gapped Domain Walls, Gapped Boundaries
  and Topological Degeneracy},''
  \href{http://dx.doi.org/10.1103/PhysRevLett.114.076402}{{\em Phys. Rev.
  Lett.} {\bfseries 114} no.~7, (2015) 076402},
  \href{http://arxiv.org/abs/1408.6514}{{\ttfamily arXiv:1408.6514
  [cond-mat.str-el]}}.

\bibitem{Kong:2019byq}
L.~Kong and H.~Zheng, ``{A mathematical theory of gapless edges of 2d
  topological orders. Part I},''
  \href{http://dx.doi.org/10.1007/JHEP02(2020)150}{{\em JHEP} {\bfseries 02}
  (2020) 150}, \href{http://arxiv.org/abs/1905.04924}{{\ttfamily
  arXiv:1905.04924 [cond-mat.str-el]}}.

\bibitem{kaidi2021higher}
J.~Kaidi, Z.~Komargodski, K.~Ohmori, S.~Seifnashri, and S.-H. Shao, ``Higher
  central charges and topological boundaries in 2+1-dimensional tqfts,'' 2021.
\newblock \url{https://arxiv.org/abs/2107.13091}.

\bibitem{davydov2013witt}
A.~Davydov, M.~M{\"u}ger, D.~Nikshych, and V.~Ostrik, ``The witt group of
  non-degenerate braided fusion categories,'' {\em Journal f{\"u}r die reine
  und angewandte Mathematik (Crelles Journal)} {\bfseries 2013} no.~677, (2013)
  135--177.

\bibitem{Fuchs:2012dt}
J.~Fuchs, C.~Schweigert, and A.~Valentino, ``{Bicategories for boundary
  conditions and for surface defects in 3-d TFT},''
  \href{http://dx.doi.org/10.1007/s00220-013-1723-0}{{\em Commun. Math. Phys.}
  {\bfseries 321} (2013) 543--575},
\href{http://arxiv.org/abs/1203.4568}{{\ttfamily arXiv:1203.4568 [hep-th]}}.

\bibitem{davydov2013structure}
A.~Davydov, D.~Nikshych, and V.~Ostrik, ``On the structure of the witt group of
  braided fusion categories,'' {\em Selecta Mathematica} {\bfseries 19} no.~1,
  (2013) 237--269, \href{http://arxiv.org/abs/1109.5558}{{\ttfamily
  arXiv:1109.5558}}.

\bibitem{Bais:2008ni}
F.~A. Bais and J.~K. Slingerland, ``{Condensate induced transitions between
  topologically ordered phases},''
  \href{http://dx.doi.org/10.1103/PhysRevB.79.045316}{{\em Phys. Rev. B}
  {\bfseries 79} (2009) 045316},
  \href{http://arxiv.org/abs/0808.0627}{{\ttfamily arXiv:0808.0627
  [cond-mat.mes-hall]}}.

\bibitem{Burnell:2017otf}
F.~J. Burnell, ``{Anyon condensation and its applications},''
  \href{http://dx.doi.org/10.1146/annurev-conmatphys-033117-054154}{{\em Ann.
  Rev. Condensed Matter Phys.} {\bfseries 9} (2018) 307--327},
  \href{http://arxiv.org/abs/1706.04940}{{\ttfamily arXiv:1706.04940
  [cond-mat.str-el]}}.

\bibitem{Ng2018higher}
S.-H. Ng, A.~Schopieray, and Y.~Wang, ``Higher gauss sums of modular
  categories,'' \href{http://dx.doi.org/10.1007/s00029-019-0499-2}{{\em Selecta
  Mathematica} {\bfseries 25} no.~4, (Aug, 2019) },
  \href{http://arxiv.org/abs/1812.11234}{{\ttfamily arXiv:1812.11234
  [math.QA]}}. \url{http://dx.doi.org/10.1007/s00029-019-0499-2}.

\bibitem{Ng2020higher}
S.-H. Ng, E.~C. Rowell, Y.~Wang, and Q.~Zhang, ``Higher central charges and
  witt groups,'' 2020.
\newblock \url{https://arxiv.org/abs/2002.03570}.

\bibitem{Kitaevanyons}
A.~Kitaev, ``{Anyons in an exactly solved model and beyond},''
  \href{http://dx.doi.org/10.1016/j.aop.2005.10.005}{{\em Annals of Physics}
  {\bfseries 321} (2006) 2--111},
  \href{http://arxiv.org/abs/cond-mat/0506438}{{\ttfamily
  arXiv:cond-mat/0506438}}.

\bibitem{KitaevE8}
A.~Kitaev, ``Toward topological classification of phases with short-range
  entanglement,'' 2011.
\newblock \url{https://online.kitp.ucsb.edu/online/topomat11/kitaev/}.

\bibitem{barkeshli2019}
M.~Barkeshli, P.~Bonderson, M.~Cheng, and Z.~Wang, ``Symmetry
  fractionalization, defects, and gauging of topological phases,''
  \href{http://dx.doi.org/10.1103/PhysRevB.100.115147}{{\em Phys. Rev. B}
  {\bfseries 100} (Sep, 2019) 115147},
  \href{http://arxiv.org/abs/arXiv:1410.4540}{{\ttfamily arXiv:1410.4540}}.

\bibitem{lapa2019}
M.~Lapa and M.~Levin, ``Anomaly indicators for topological orders with u(1) and
  time-reversal symmetry,''
  \href{http://dx.doi.org/10.1103/physrevb.100.165129}{{\em Physical Review B}
  {\bfseries 100} no.~16, (Oct, 2019) },
  \href{http://arxiv.org/abs/arXiv:1905.00435}{{\ttfamily arXiv:1905.00435}}.
  \url{http://dx.doi.org/10.1103/PhysRevB.100.165129}.

\bibitem{benini2019}
F.~Benini, C.~C{\'o}rdova, and P.-S. Hsin, ``On 2-group global symmetries and
  their anomalies,'' \href{http://dx.doi.org/0.1007/JHEP03(2019)118}{{\em
  Journal of High Energy Physics} {\bfseries 2019} no.~3, (Mar, 2019) 118},
  \href{http://arxiv.org/abs/arXiv:1803.09336}{{\ttfamily arXiv:1803.09336}}.

\bibitem{kobayashi2021spinc}
R.~Kobayashi and M.~Barkeshli, ``(3+1)d path integral state sums on curved u(1)
  bundles and u(1) anomalies of (2+1)d topological phases,'' 2021.
\newblock \url{https://arxiv.org/abs/2111.14827}.

\bibitem{Seiberg2016Gapped}
N.~Seiberg and E.~Witten, ``{Gapped boundary phases of topological insulators
  via weak coupling},'' \href{http://dx.doi.org/10.1093/ptep/ptw083}{{\em
  Progress of Theoretical and Experimental Physics} {\bfseries 2016} no.~12,
  (2016) 1--86}, \href{http://arxiv.org/abs/arXiv:1602.04251v3}{{\ttfamily
  arXiv:1602.04251v3}}.

\bibitem{Thorngren2021end}
R.~Thorngren and Y.~Wang, ``Anomalous symmetries end at the boundary,''
  \href{http://dx.doi.org/10.1007/JHEP09(2021)017}{{\em Journal of High Energy
  Physics} no.~9, (2021) 17}, \href{http://arxiv.org/abs/2012.15861}{{\ttfamily
  arXiv:2012.15861}}.

\bibitem{chen2012spt}
X.~Chen, Z.-C. Gu, Z.-X. Liu, and X.-G. Wen, ``Symmetry-protected topological
  orders in interacting bosonic systems,''
  \href{http://dx.doi.org/10.1126/science.1227224}{{\em Science} {\bfseries
  338} no.~6114, (2012) 1604--1606},
  \href{http://arxiv.org/abs/http://www.sciencemag.org/content/338/6114/1604.full.pdf}{{\ttfamily
  http://www.sciencemag.org/content/338/6114/1604.full.pdf}}.

\bibitem{bulmashSymmFrac}
D.~Bulmash and M.~Barkeshli, ``Fermionic symmetry fractionalization in
  (2+1)d,'' 2021.
\newblock \url{https://arxiv.org/abs/2109.10913}.

\bibitem{cheng2016lsm}
M.~Cheng, M.~Zaletel, M.~Barkeshli, A.~Vishwanath, and P.~Bonderson,
  ``Translational symmetry and microscopic constraints on symmetry-enriched
  topological phases: A view from the surface,''
  \href{http://dx.doi.org/10.1103/PhysRevX.6.041068}{{\em Phys. Rev. X}
  {\bfseries 6} (Dec, 2016) 041068},
  \href{http://arxiv.org/abs/arXiv:1511.02263}{{\ttfamily arXiv:1511.02263}}.
  \url{https://link.aps.org/doi/10.1103/PhysRevX.6.041068}.

\bibitem{Reshetikhin1991}
N.~Reshetikhin and V.~G. Turaev, ``Invariants of 3-manifolds via link
  polynomials and quantum groups,''
  \href{http://dx.doi.org/10.1007/BF01239527}{{\em Inventiones mathematicae}
  {\bfseries 103} no.~1, (1991) 547--597}.
  \url{https://doi.org/10.1007/BF01239527}.

\bibitem{Turaev+2010}
V.~G. Turaev, \href{http://dx.doi.org/doi:10.1515/9783110221848}{{\em Quantum
  Invariants of Knots and 3-Manifolds}}.
\newblock De Gruyter, 2010.
\newblock \url{https://doi.org/10.1515/9783110221848}.

\bibitem{drinfeld2010braided}
V.~Drinfeld, S.~Gelaki, D.~Nikshych, and V.~Ostrik, ``On braided fusion
  categories i,'' 2010.
\newblock \url{https://arxiv.org/abs/0906.0620}.

\bibitem{delmastro2021}
D.~Delmastro, D.~Gaiotto, and J.~Gomis, ``Global anomalies on the hilbert
  space,'' 2021.
\newblock \url{http://arxiv.org/abs/2101.02218}.

\bibitem{aasen21ferm}
D.~Aasen, P.~Bonderson, and C.~Knapp, ``Characterization and classification of
  fermionic symmetry enriched topological phases,'' 2021.
\newblock \url{https://arxiv.org/abs/2109.10911}.

\bibitem{gu2014}
Z.-C. Gu and X.-G. Wen, ``Symmetry-protected topological orders for interacting
  fermions: Fermionic topological nonlinear $\ensuremath{\sigma}$ models and a
  special group supercohomology theory,''
  \href{http://dx.doi.org/10.1103/PhysRevB.90.115141}{{\em Phys. Rev. B}
  {\bfseries 90} (Sep, 2014) 115141},
  \href{http://arxiv.org/abs/arXiv:1201.2648}{{\ttfamily arXiv:1201.2648}}.
  \url{https://link.aps.org/doi/10.1103/PhysRevB.90.115141}.

\bibitem{Kapustin:2014dxa}
A.~Kapustin, R.~Thorngren, A.~Turzillo, and Z.~Wang, ``{Fermionic Symmetry
  Protected Topological Phases and Cobordisms},''
  \href{http://dx.doi.org/10.1007/JHEP12(2015)052}{{\em JHEP} {\bfseries 12}
  (2015) 052},
\href{http://arxiv.org/abs/1406.7329}{{\ttfamily arXiv:1406.7329
  [cond-mat.str-el]}}.

\bibitem{qingrui}
Q.-R. Wang and Z.-C. Gu, ``Construction and classification of symmetry
  protected topological phases in interacting fermion systems,''
  \href{http://dx.doi.org/10.1103/PhysRevX.10.031055}{{\em Physical Review X}
  {\bfseries 10} (2018) 031055},
  \href{http://arxiv.org/abs/1811.00536}{{\ttfamily arXiv:1811.00536}}.

\bibitem{Thorngren2018bosonization}
R.~Thorngren, ``Anomalies and bosonization,''
  \href{http://arxiv.org/abs/arXiv:1810.04414}{{\ttfamily arXiv:1810.04414}}.
  \url{https://arxiv.org/abs/1810.04414}.

\bibitem{BCHM2021classification}
M.~Barkeshli, Y.-A. Chen, P.-S. Hsin, and N.~Manjunath, ``Classification of
  (2+1)d invertible fermionic topological phases with symmetry,'' 2021.
\newblock \url{https://arxiv.org/abs/2109.11039}.

\bibitem{bhardwajGaiottoKapustin2017}
L.~Bhardwaj, D.~Gaiotto, and A.~Kapustin, ``State sum constructions of
  spin-tfts and string net constructions of fermionic phases of matter,''
  \href{http://dx.doi.org/10.1007/JHEP04(2017)096}{{\em Journal of High Energy
  Physics} {\bfseries 2017} no.~96, (2017) }.
  \url{https://doi.org/10.1007/JHEP04(2017)096}.

\bibitem{ChenjieWang2017}
Y.~Wan and C.~Wang, ``Fermion condensation and gapped domain walls in
  topological orders,'' \href{http://dx.doi.org/10.1007/jhep03(2017)172}{{\em
  Journal of High Energy Physics} {\bfseries 2017} no.~3, (Mar, 2017) },
  \href{http://arxiv.org/abs/1607.01388}{{\ttfamily arXiv:1607.01388
  [cond-mat.str-el]}}. \url{http://dx.doi.org/10.1007/JHEP03(2017)172}.

\bibitem{Heinrich2018}
C.~Heinrich and M.~Levin, ``Criteria for protected edge modes with
  $\mathbb{Z}_2$ symmetry,''
  \href{http://dx.doi.org/10.1103/physrevb.98.035101}{{\em Physical Review B}
  {\bfseries 98} no.~3, (Jul, 2018) },
  \href{http://arxiv.org/abs/1803.00023}{{\ttfamily arXiv:1803.00023
  [cond-mat]}}.

\bibitem{TJF2021minimal}
T.-J. Freyd and D.~Reutter, ``Minimal nondegenerate extensions,'' 2021.
\newblock \url{https://arxiv.org/abs/2105.15167}.

\bibitem{bruillard2017a}
P.~Bruillard, C.~Galindo, T.~Hagge, S.-H. Ng, J.~Y. Plavnik, E.~C. Rowell, and
  Z.~Wang, ``Fermionic modular categories and the 16-fold way,''
  \href{http://dx.doi.org/10.1063/1.4982048}{{\em Journal of Mathematical
  Physics} {\bfseries 58} no.~4, (Apr, 2017) 041704},
  \href{http://arxiv.org/abs/arXiv:1603.09294}{{\ttfamily arXiv:1603.09294}}.
  \url{http://dx.doi.org/10.1063/1.4982048}.

\bibitem{Tachikawa:2017gyf}
Y.~Tachikawa, ``{On gauging finite subgroups},''
  \href{http://dx.doi.org/10.21468/SciPostPhys.8.1.015}{{\em SciPost Phys.}
  {\bfseries 8} no.~1, (2020) 015},
  \href{http://arxiv.org/abs/1712.09542}{{\ttfamily arXiv:1712.09542
  [hep-th]}}.

\bibitem{tata2021anomalies}
S.~Tata, R.~Kobayashi, D.~Bulmash, and M.~Barkeshli, ``Anomalies in (2+1)d
  fermionic topological phases and (3+1)d path integral state sums for
  fermionic spts,'' 2021.
\newblock \url{https://arxiv.org/abs/2104.14567}.

\bibitem{thorngren2015framed}
R.~Thorngren, ``Framed wilson operators, fermionic strings, and gravitational
  anomaly in 4d,'' \href{http://dx.doi.org/10.1007/jhep02(2015)152}{{\em
  Journal of High Energy Physics} {\bfseries 2015} no.~2, (Feb, 2015) }.
  \url{http://dx.doi.org/10.1007/JHEP02(2015)152}.

\bibitem{Gaiotto:2015zta}
D.~Gaiotto and A.~Kapustin, ``{Spin TQFTs and Fermionic Phases of Matter},''
  \href{http://dx.doi.org/10.1142/S0217751X16450445}{{\em Int. J. Mod. Phys.}
  {\bfseries A31} no.~28n29, (2016) 1645044},
\href{http://arxiv.org/abs/1505.05856}{{\ttfamily arXiv:1505.05856
  [cond-mat.str-el]}}.

\bibitem{aasen2019}
D.~Aasen, E.~Lake, and K.~Walker, ``Fermion condensation and super pivotal
  categories,'' \href{http://dx.doi.org/10.1063/1.5045669}{{\em Journal of
  Mathematical Physics} {\bfseries 60} no.~12, (Dec, 2019) 121901},
  \href{http://arxiv.org/abs/arXiv:1709.01941}{{\ttfamily arXiv:1709.01941}}.
  \url{http://dx.doi.org/10.1063/1.5045669}.

\bibitem{ManifoldAtlasWu}
{Karlheinz Knapp}, ``Wu class.''
\newblock \url{http://www.map.mpim-bonn.mpg.de/Wu_class}.

\bibitem{tata2020}
S.~Tata, ``Geometrically interpreting higher cup products, and application to
  combinatorial pin structures,'' 2020.
\newblock \url{https://arxiv.org/abs/2008.10170}.

\end{thebibliography}\endgroup
\end{document}